\documentclass[11pt,a4paper,reqno]{amsart}

\usepackage{xcolor}
\usepackage{csquotes}
\usepackage{amsmath}
\usepackage{csquotes} %enquote env
\usepackage{siunitx} %SI{}{} env
\usepackage{upgreek}
\usepackage{amssymb}
\usepackage{graphicx}
\usepackage{subcaption}
\usepackage[headings]{fullpage} %textwidth
\usepackage{setspace} %linespace
\renewcommand{\baselinestretch}{1.1} 
\usepackage{booktabs}
\usepackage{comment}

\usepackage{lineno} %line numbering

\usepackage[square,numbers,sort&compress]{natbib}

\usepackage[foot]{amsaddr} % Affiliations on first page
\definecolor{ulmgruen}{rgb}{0.3372,0.6667,0.1098}
\definecolor{pythonblue}{HTML}{1F77B4}
\definecolor{pythonorange}{HTML}{ff7f0e}

\DeclareMathOperator*{\argmax}{argmax}
\DeclareMathOperator*{\argmin}{argmin}

\DeclareMathOperator*{\aed}{aed}
\DeclareMathOperator*{\mae}{error}
\DeclareMathOperator*{\dist}{dist}

\DeclareMathOperator{\EX}{\mathbb{E}}
\newcommand{\R}{\mathbb{R}}
\newcommand{\Z}{\mathbb{Z}_\rho}

\newcommand{\N}{\mathbb{N}}
\newcommand{\G}{\mathcal{G}}

\newcommand{\tpp}{\text{prob}}
\newcommand{\TPP}{\text{prob}}
\newcommand{\distToBack}{{\text{prob}_\text{dist}}}
\newcommand{\crackSize}{{\text{prob}_\text{size}}}
\newcommand{\diam}{\text{diam}}

\newcommand{\avTPP}{\xoverline{\text{prob}}} % function
\newcommand{\avdistToBack}{\xoverline{\text{prob}}_\text{dist}}
\newcommand{\avcrackSize}{\xoverline{\text{prob}}_\text{size}}

\newcommand{\ssa}{\sigma}

\newcommand{\Long}{_\text{long}}
\newcommand{\Short}{_\text{short}}
\newcommand{\modelShort}{P_{\widehat{\theta}_{\Short}}}
\newcommand{\modelLong}{P_{\widehat{\theta}_{\Long}}}

\newcommand{\LongUp}{^{(2)}}
\newcommand{\ShortUp}{^{(1)}}
\newcommand{\exUp}{\text{(ex)}}
\newcommand{\prUp}{\text{(pr)}}
\newcommand{\thetaUp}{\text{($\theta$)}}

\newcommand{\skel}{\mathcal{S}}
\newcommand{\WeiInd}{_\text{W}}
\newcommand{\PoiInd}{_\text{P}}
\newcommand{\DimInd}{_\text{dim}}
\newcommand{\thetaBar}{\theta_1}
\newcommand{\nCracks}{n_\text{cracks}}
\newcommand{\NCracks}{N_\text{cracks}}
\newcommand{\nFacets}{N_\text{facets}}

\newcommand{\solid}{\Xi_\solidText}
\newcommand{\crack}{\Xi_\crackText}
\newcommand{\BG}{\Xi_\text{BG}}
\newcommand{\solidText}{\text{solid}}
\newcommand{\crackText}{\text{crack}}
\newcommand\nubig[1]{\nu_3\bigl(#1\bigr)}

\newcommand{\startingSet}{{H_\text{S}}}
\newcommand{\targetSet}{{H_\text{T}}}

\newcommand{\transportPhase}{{\Xi}}

\newcommand{\LE}{{\tau_\text{LE}}}
\newcommand{\SE}{{\tau_\text{SE}}}

\newcommand\restr[2]{{% we make the whole thing an ordinary symbol
  \left.\kern-\nulldelimiterspace % automatically resize the bar with \right
  #1 % the function
  \vphantom{\big|} % pretend it's a little taller at normal size
  \right|_{#2} % this is the delimiter
  }}

\makeatletter
\newsavebox\myboxA
\newsavebox\myboxB
\newlength\mylenA

\newcommand*\xoverline[2][0.75]{%
    \sbox{\myboxA}{$\m@th#2$}%
    \setbox\myboxB\null% Phantom box
    \ht\myboxB=\ht\myboxA%
    \dp\myboxB=\dp\myboxA%
    \wd\myboxB=#1\wd\myboxA% Scale phantom
    \sbox\myboxB{$\m@th\overline{\copy\myboxB}$}%  Overlined phantom
    \setlength\mylenA{\the\wd\myboxA}%   calc width diff
    \addtolength\mylenA{-\the\wd\myboxB}%
    \ifdim\wd\myboxB<\wd\myboxA%
       \rlap{\hskip 0.5\mylenA\usebox\myboxB}{\usebox\myboxA}%
    \else
        \hskip -0.5\mylenA\rlap{\usebox\myboxA}{\hskip 0.5\mylenA\usebox\myboxB}%
    \fi}
\makeatother
\patchcmd{\subsubsection}{\itshape}{\bfseries}{}{}%Bold subsubsections
\patchcmd{\paragraph}{\normalfont}{\bfseries}{}{}%Bold paragraphs

\begin{document}

	\author[]{Philipp Rieder$^1$,
		Orkun Furat$^1$,
		Francois L.E. Usseglio-Viretta$^2$,
		Jeffery Allen$^2$,
		Peter J. Weddle$^2$,
		Donal P. Finegan$^2$,
		Kandler Smith$^2$,
		Volker Schmidt$^1$}
	\address{$^1$Institute of Stochastics, Ulm University, Helmholtzstra{\upshape{\ss}}e~18,89069~Ulm, Germany}
	\address{$^2$National Renewable Energy Laboratory, 15013 Denver W Parkway, Golden, CO 80401, USA}

	\title[Reconstruction of cracked  polycrystalline NMC particles]{Stochastic 3D reconstruction of cracked polycrystalline NMC particles using 2D SEM data}

	\begin{abstract}
        Li-ion battery performance is strongly influenced by their cathodes' properties and consequently by the 3D microstructure of the particles the cathodes are comprised of. During calendaring and cycling, cracks develop within cathode particles, which may affect  performance in multiple ways. On the one hand, cracks reduce internal connectivity  such that  electron transport within cathode particles is hindered. On the other hand, intra-particle cracks can increase the cathode reactive surface. Due to these contradictory effects, it is necessary to quantitatively investigate how battery cycling effects cracking and how cracking in-turn influences battery performance. Thus, it is necessary to characterize the 3D particle morphology with structural descriptors and quantitatively correlate them with effective battery properties. Typically, 3D structural characterization is performed using image data. However, informative 3D imaging techniques are time-consuming, costly and rarely available, such that analyses often have to rely on 2D image data. This paper presents a novel stereological approach for generating virtual 3D cathode particles that exhibit crack networks that are statistically equivalent to those observed in 2D sections of experimentally measured particles. Consequently,  more easily available 2D image data suffices for deriving a full 3D characterization of cracked cathodes particles. In future research, the virtually generated 3D particles will be used as geometry input for spatially resolved electro-chemo-mechanical simulations, to enhance our understanding of structure-property relationships of cathodes in Li-ion batteries.
	\end{abstract}
	
	\maketitle
	\noindent\textit{Key words and phrases.} 
  Lithium-ion battery, cathode, active material, NMC particle, crack morphology, stereological 3D reconstruction, stochastic crack network model%, digital twin
	\section{Introduction} \label{sec:intro}

%------------\paragraph{Revised introduction}---------------
Li-ion batteries are vital to modern technology and transportation~\cite{EP10,MACA16}.  Current research initiatives in Li-ion technology aim to increase battery energy density while simultaneously extending cycle-life~\cite{HMLLLZMGCGBCSD19,CDTPJS19}.  
To that end, high-voltage LiNi$_x$Mn$_y$Co$_z$O$_2$ (NMC$xyz$) 
 cathodes are increasingly integrated into energy-dense cells due to their high average voltages ($\approx$ \SI{3.7}{\volt} vs.~Li) and high theoretical specific capacities  ($185-\SI{278}{\milli\ampere}$h~g$^{-1}$)~\cite{LCRUPAW11,MSL17,JHHM18,LLXCNBD17,LYZSSB19}.  
Additionally, when cycled within appropriate voltage windows ($\leq$ \SI{4.3}{\volt} vs.~Li), these cathodes can reach upwards of a thousand cycles with high capacity retention~\cite{HMLLLZMGCGBCSD19}. 
Currently, NMC chemistries are the primary candidates for cathode materials that lead to energy-dense Li-ion batteries, spanning both liquid~\cite{cui2020understanding,TYCCGSWDBSDEDTPJ21,TWYCCFLPKAUCBDSCWWR22,DABCCCGKKMRRSTUW22,LBCDGKLLLMMSTVWXXYYZ19} and solid-state~\cite{Wang2017Quantitative,WYHLLZSWHLHZZLSS20,TWHBK23,CRSTFRJ21,KALDZBHZJ17} electrolyte systems. 
However, NMC cathodes can exhibit capacity-fade mechanisms including transition-metal dissolution~\cite{YXZLMLWSXWKYDYXLPCLZLL19,L20}, surface reconstruction~\cite{LLXCNBD17,cui2020understanding}, electrolyte reactivity and gassing~\cite{LLXCNBD17,JMMSG17,JSMSG18}, and particle cracking~\cite{XSVZ17,RK18}.  These fade mechanisms are highly coupled~\cite{TWYCCFLPKAUCBDSCWWR22,VNWVBWWVH05,LYZSSB19,LSMCM23,TYFCLWBCDWTESADQDTJ22,RPYS18,MXCZHMLL24,Wang2017Quantitative}.  
For example, in liquid systems, particle cracking can expose uncoated active material to liquid electrolyte, resulting in increased electrochemical side reactions and surface reconstruction~\cite{LLXCNBD17,LSMCM23}.  In solid systems, cathode cracking can result in reduced surface area to either the solid-state electrolyte or the electron-conducting phase, resulting in increased charge transfer and ohmic resistance, respectively~\cite{TWHBK23,CRSTFRJ21,KALDZBHZJ17}. 

Because a significant amount of cathode capacity-fade mechanisms are related to secondary-particle cracking, researchers typically evaluate cathode ``aging'' through qualitative crack analysis~\cite{TYFCLWBCDWTESADQDTJ22,TWYCCFLPKAUCBDSCWWR22}.  Cathode particle cracking can occur for different reasons.  First, during manufacturing, cathode cracking can occur during the calendaring process~\cite{JHHM18,XPSSWDF23,BINGK16}.  The cracks formed during calendaring typically originate at particle/particle or particle/current-collector contacts and tend to form long cracks that cleave  particles.  Second, cathode cracking can occur during formation cycles due to non-ideal primary particle grain orientations.  
These initial ``break-in'' cracks tend to be small and are significantly influenced by the grain shapes, sizes, and orientations~\cite{AWVMUFCMSFDTS21,KLCYPC18}.  Break-in cracking is currently the primary focus for physics-based chemo-mechanical models~\cite{AWVMUFCMSFDTS21,TWBK21,BZLSX20,SSLMXFZ24,CARFSSWSX24}. Finally, cracks can form during operation when the cathodes are cycled at higher voltage ranges, either due to increased voltage bounds or due to voltage slippage~\cite{DXGD20,TWYCCFLPKAUCBDSCWWR22}.  At high voltages or during high delithiation demands, the lithium concentration on the cathode surface can drop below a minimum concentration threshold causing irreversible reconstruction of the host crystal.  This reconstruction reduces the specific capacity and induces significant local stain, leading to secondary-particle cracking~\cite{AWVMUFCMSFDTS21,RPYS18}.

Currently, structural post-mortem analysis of cathode particle fracture is primarily conducted using 2D scanning electron microscope (SEM) images and X-ray techniques~\cite{TYCCGSWDBSDEDTPJ21,TWYCCFLPKAUCBDSCWWR22,FFYTSS22,FURAT2024,JKKJYJS20,RPYS18}.  
Since a quantitative analysis of such 2D images can be difficult, the comparison of differently aged post-mortem cathodes is often performed by means of visual inspection~\cite{TYCCGSWDBSDEDTPJ21,TWYCCFLPKAUCBDSCWWR22,MXCZHMLL24}.  
Such  a qualitative analysis is typically accompanied by quantitative electrochemical analysis (e.g., electrochemical impedance spectroscopy, incremental capacity analysis~\cite{TWYCCFLPKAUCBDSCWWR22,RPYS18}) and post-mortem, atomistic-scale surface-sensitive techniques~\cite{TYCCGSWDBSDEDTPJ21,JGHPSKHYK13,JKKJYJS20,YXZLMLWSXWKYDYXLPCLZLL19,RPYS18}.  However, relying on qualitative cracking-extent assessments introduces subjectivity in the analysis, highlighting the need for more quantitative and reproducible methods to characterize cathode-particle fracture.   
A quantitative analysis of cracks in 2D SEM data has been conducted, e.g., in \cite{FFYTSS22,FURAT2024}. However,  2D images of cracked particles depict only planar sections of the actual 3D microstructure.  In other words, a 2D slice of a cathode electrode represents just a small  portion of the 3D system, which includes out-of-plane features such as tortuous crack connections. 

In contrast to 2D crack analysis, it is significantly more difficult to segment and identify crack structures in 3D images \cite{PETRICH2017,westhoff2018} and to reassemble fragments of fractured particles \cite{WILHELM2024}. 
This increased difficulty is due to the fact that 3D imaging (e.g., via nano-CT) is often accompanied with a lower resolution than 2D microscopy techniques (e.g., SEM), which produce image data on a similar length scale---i.e., fine structures caused by cracks often exhibit a bad contrast in 3D image data.
Moreover, a quantitative crack analysis requires computation of metrics to describe cracks in 3D~\cite{XMXWWLPZYLL18}.
Unfortunately, the necessary 3D imaging equipment is expensive and often less available than comparable 2D imaging equipments and their analysis tools~\cite{WLHTBJS23,SZJZ20,YXZLMLWSXWKYDYXLPCLZLL19}.
A potential remedy is provided by stochastic 3D modeling, which can generate countless virtual NMC particles exhibiting statistically similar properties as the relatively low number of particles that have been imaged in 3D~\cite{Furat2021}.
Realizations of these stochastic 3D models can serve as geometry inputs for mechanical and electrochemical simulations to investigate crack propagation in NMC particles~\cite{AWVMUFCMSFDTS21,HAN2024} or more generally polycrystalline media~\cite{Willot2020}. By performing such simulation studies on generated morphologies  quantitative structure-property relationships can be derived \cite{prifling2021large,neumann2017microstructure}.

As mentioned above, measured 3D image data is not always accessible. Therefore approaches have been developed to calibrate stochastic 3D models utilizing only 2D image data \cite{liebscher2015stereological}. Recently, a stochastic nanostructure model based on generative adversarial networks (GAN) was introduced, which mimics the 3D polycrystalline grain architecture of non-cracked NMC particles~\cite{fuchs2024} by only using 2D electron backscatter diffraction (EBSD) cross sections for calibration.  In the present paper, a stochastic 3D model is proposed that can generate realistic cracks in virtual polycrystalline NMC particles, which propagate along grain boundaries. Similar to the approach considered in \cite{jung2023}, our model is based on random  tessellations, where certain facets are dilated to mimic cracks. In this work, a facet between two grains is either intact or fully cracked without intermediate case (that is all surface elements of the facet are either intact or cracked. The model is calibrated and validated by comparing  planar cross sections of the  stochastic 3D crack network model with 2D SEM image data, utilizing several geometric descriptors characterizing the morphology of the crack phases. Additionally, to emphasize the strength of our stereological modeling approach, geometric descriptors related to effective battery properties are determined, which cannot be reliably derived from 2D images.

\section{Materials and image processing}\label{sec:Materials}

The focus of this section is to describe the  materials considered in the present paper, as well as on the processing of 2D SEM image data of these materials. First, in Section \ref{ssec:Aging}, the cathode materials and their cycling history are discussed. Then, in Section \ref{ssec:imageProcessing}, several  image processing techniques are described, where gray-scale images of planar cross sections of the cathodes, obtained by SEM imaging, are phasewise segmented using a 2D U-net and, afterwards,  particlewise segmented utilizing a marker-based watershed transform. Additionally, morphological operations are used to denoise the crack phase. Finally, in Section \ref{ssec:Subdivison}, the set of segmented 2D images is decomposed into two subsets, based on the predominance of short or long cracks.

\subsection{Electrode materials and cycling history}\label{ssec:Aging}
The active electrode material used in the present paper consisted of LiNi$_{0.5}$Mn$_{0.3}$Co$_{0.2}$O$_{2}$ (NMC532) and was taken from the same batch of cells cycled in our previous work \cite{FURAT2024}, where the particles had similar polycrystalline architectures as those shown in~\cite{Quinn2020}. The electrodes consisted of 90 wt\% NMC532, 5wt\% Timcal C45 carbon and 5wt\% Solvay 5130 PVDF binder. The coating thickness was \SI{62}{~\micro\meter} with 26.1\% porosity.

The cell was formed by charging to \SI{1.5}{\volt}, holding at open-circuit for 12 hours, and then cycling 3 times between \SI{3}{\volt} and \SI{4.1}{\volt} using a protocol consisting of C/2 constant-current and constant-voltage at \SI{4.1}{\volt} until the current dropped  below C/10. The cells were then degassed, resealed, and prepared for fast charging at $\SI{20}{\celsius}$. Subsequently, the cells were cycled using a protocol of fast charging at 6C constant current between \SI{3}{\volt} and  \SI{4.1}{\volt} followed by constant-voltage hold until 10 minutes total charge time had elapsed. Charge was followed by 15 minutes of open circuit and discharge at C/2 to \SI{3}{\volt}, followed by a final rest for 15 minutes. The materials used in this paper were cycled 200 times under these conditions.

\subsection{ Preprocessing and analysis of 2D SEM image data}\label{ssec:imageProcessing}
\begin{figure}[h]
     \centering
      \subcaptionbox*{}{\frame{\includegraphics[width=.49\textwidth]{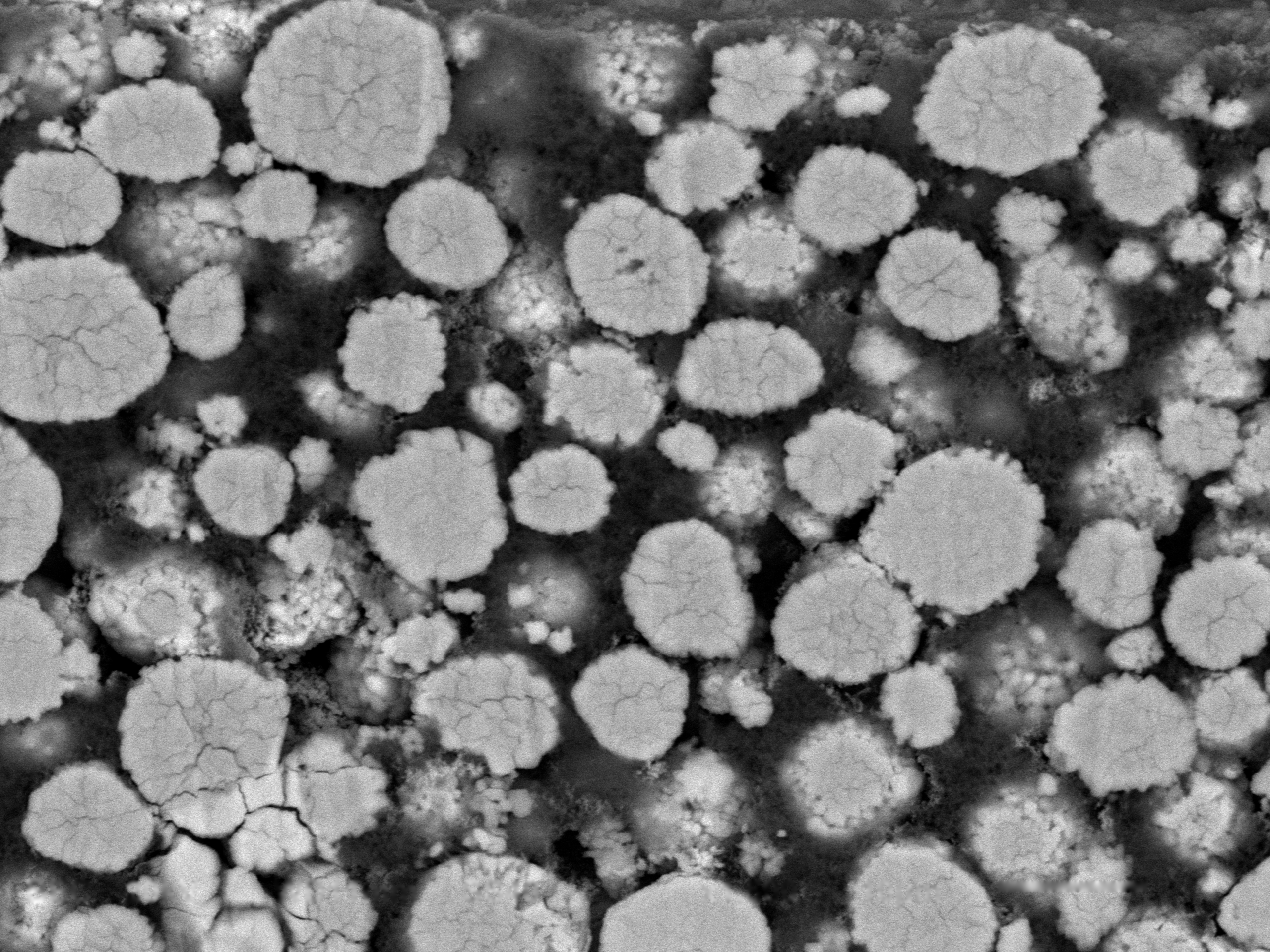}}}
    \subcaptionbox*{}{\frame{\includegraphics[width=.49\textwidth]{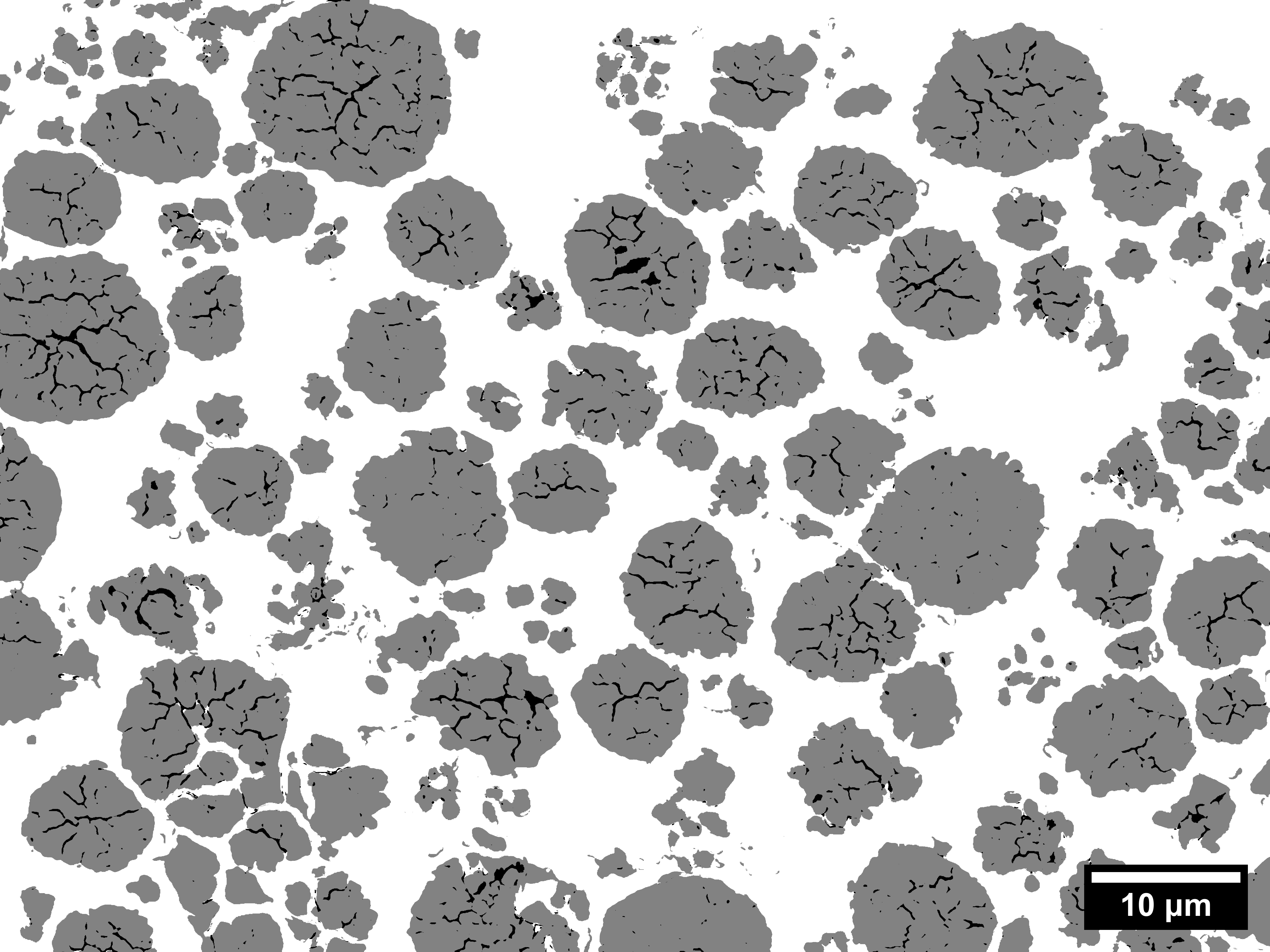}}}
     \caption{2D SEM image (left) and its phasewise segmentation (right), where each pixel is classified as background (white), solid (gray) or crack (black). }
     \label{fig:SEMCrossSection}
 \end{figure}

% resolution model 36 nm;
% resolution experimental=14.29 nm; 
The NMC electrode material was removed from the cells and a small sample cut from the electrode sheet. The sample was then cross sectioned using an Ar-ion beam cross-sectional polisher (JOEL CP19520). The cross-sectioned face was then imaged in an SEM system with a pixel size of \SI{14.29}{~\nano\meter}. A representative cross section, derived by SEM imaging, is presented in Figure~\ref{fig:SEMCrossSection} (left).

For image processing, we first describe the image processing steps which were performed in \cite{FURAT2024} to segment the 2D SEM image data of the cross sections with respect to phases and particles, i.e., each pixel is classified either as solid phase, crack phase or background, where each particle is assigned with a unique label. 
Note that the raw image data depicted scale bars for indicating the corresponding length scales (which have been produced by the microscope's software). Since the scale bars can adversely impact subsequent image processing steps, the \textit{inpainting\_biharmonic} method of the scikit-image package in Python \cite{VanderWalt2014} has been utilized to remove scale bars. Then, a generative adversarial network \cite{furat2022super}  has been deployed to increase the resolution (super-resolution) of  image data such that the assignment of pixels to phases (i.e., solid phase, crack phase, background) can be performed more reliably.

To obtain a phasewise segmentation, a 2D U-net was deployed to classify the phase affiliation for each pixel in 2D SEM cross sections. More precisely, the network's output is given by pixelwise probabilities of phase membership. By deploying thresholding techniques onto these pixelwise probabilities, a phase-wise segmentation has been obtained, see \cite{FURAT2024} for further details. An exemplary phasewise segmented cross section is shown in Figure \ref{fig:SEMCrossSection} (right). In particular, Figure~\ref{fig:SEMCrossSection} indicates that the data has been segmented reasonably well, i.e., only a low, statistically negligible number of particles (see bottom left) exhibits larger misclassified areas.

The particle-wise segmentation was obtained by means of a marker-based watershed transform on the Euclidean distance transform, denoted by $D$ in the following. 
More precisely, $D\colon W \to \R_+=[0,\infty)$ is a mapping, which assigns each pixel $x\in W$ its distance to the background phase. Here, $W\subset \Z^2$ represents the sampling window, where $\Z=\{\ldots,-\rho,0,\rho,\ldots\}$ and $\rho>0$ denotes the pixel length. Note that  the watershed function of the Python package scikit-image \cite{VanderWalt2014} was deployed on $-D$, where the markers (i.e., the positions of particles to be segmented) are obtained by thresholding $D$ at some distance level $r>0$, where $r$ is set equal to 50 pixels.
After the application of the watershed algorithm, truncated particles were removed in order to avoid edge effects.
This procedure is performed on 13 SEM cross sections, each consisting of $5973\times 3079$ pixels, which corresponds to approximately $\SI{85}{\micro \meter} \times \SI{44}{\micro \meter}$ with a resolution of $\rho=\SI{14.29}{\nano\meter}$. Note that these 13 cross sections are derived from the same cathode and are partly overlapping. In each cross section, between 43 and 60 particles were detected. 

% ---------Infos about cross sections-------

In addition to the preprocessing procedure explained above and proposed in \cite{FURAT2024}, the following data processing steps have been carried out. First, since some SEM images depict overlapping areas, duplicates were removed. More precisely, all pairs of particles from overlapping images were registered, i.e., for each pair of particles a rigid transformation is determined which maximizes the correspondence of the first particle after application of the transformation with the second particle, where the agreement is measured by means of the cross correlation in scikit-image \cite{VanderWalt2014}.
If pairs of registered particles exhibit a large correspondence, a duplicate is detected, which is omitted in further analysis. 
Furthermore, to reduce the number of very small cracks, e.g., caused by noise or by several connected components of the crack phase that actually belong to the same crack, morphological opening, followed by morphological closing, was performed on the crack phase. For both morphological operations,  a disk-shaped structuring element with radius $r_o=1$ for opening and $r_c=3$ for closing was used. 

In summary, the image processing  procedure described above resulted into 506 images depicting the phasewise segmentation of  NMC particles in planar sections, i.e., each pixel is classified either as solid (active NMC material), crack, or background. 
Each of these images depicts a single cross section of a NMC particle, which shows a certain network of cracks.  In the following sections, an individual particle is denoted by $P_\text{ex}$.

\subsection{Decomposition of the set of segmented particles into two subsets  
}\label{ssec:Subdivison}

In this section we explain how the set of segmented particles, described in Section \ref{ssec:imageProcessing}, is decomposed into two subsets with predominantly long and short cracks, respectively. For this classification, for each particle $P_\text{ex}$,  we  consider a continuous representation in the two-dimensional Euclidean space $\R^2$, denoting its solid  phase by $\solid^\exUp\subset\R^2$ and its crack phase by $\crack^\exUp\subset\R^2$, where each pixel of $P_\text{ex}$ is considered as patch (i.e., as a square subset of $\R^2$). 
Thus, in the following we will write 
\begin{equation}\label{for.con.rep}
P_\text{ex}=\bigl(\solid^\exUp,\crack^\exUp\bigr)
\end{equation}
for the continuous representation of a particle. Furthermore,  by $\G$ we denote the set of continuous representations of all 506 particles. 
The dataset $\G$ is comprised of particles with sizes ranging from \SI{1.39}{\micro \meter} to \SI{13.62}{\micro \meter} (in terms of their area-equivalent diameters, denoted by $\aed(P_{\text{ex}})$).

\begin{figure}[h]
     \centering
    \subcaptionbox*{}{\includegraphics[scale=.2]{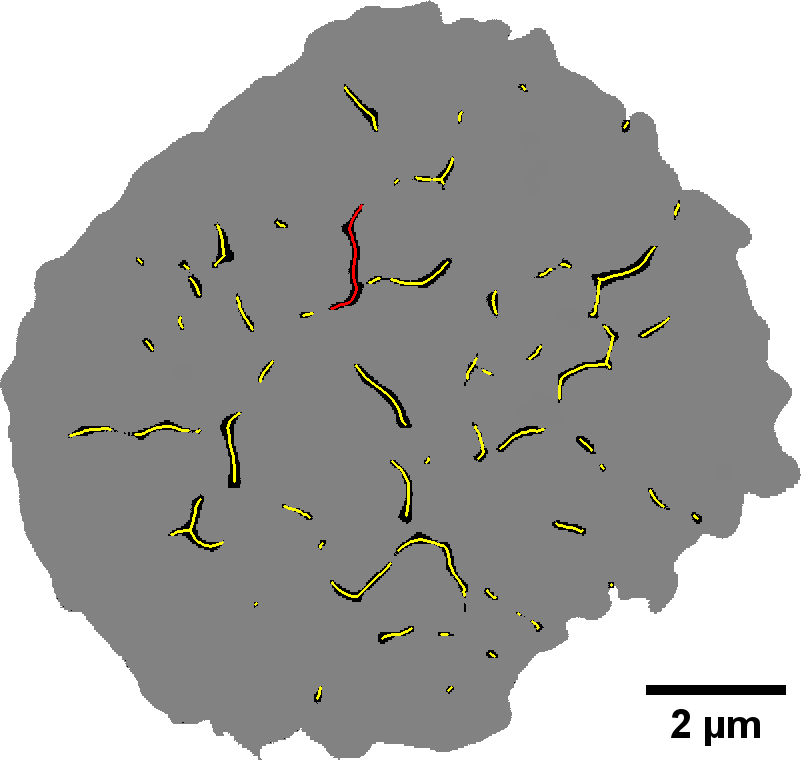}}
    \subcaptionbox*{}{\includegraphics[scale=.2]{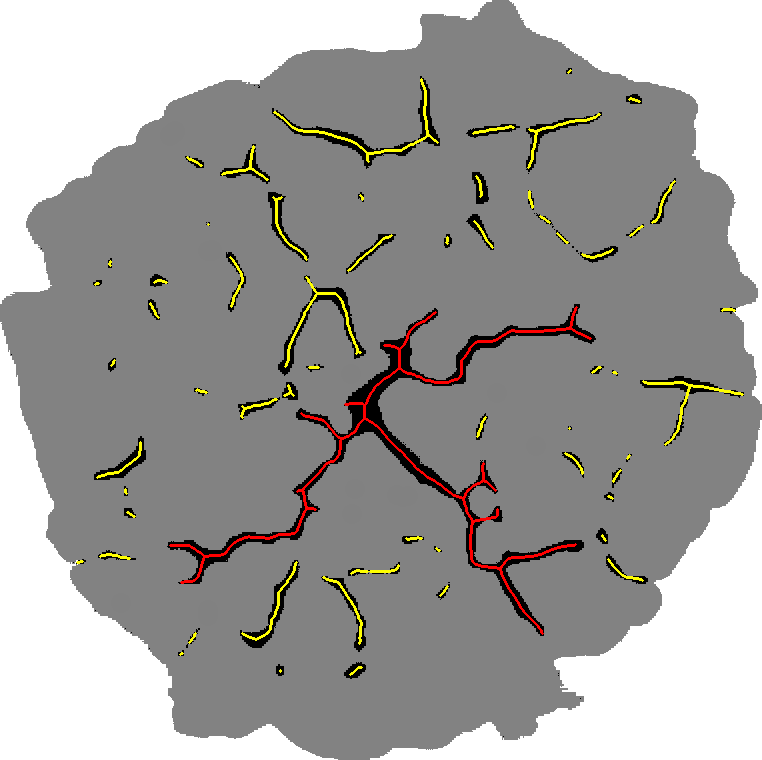}}
     \caption{
     Segmented NMC particles together with their skeletons (yellow), where the skeleton segment of the longest crack is highlighted in red.
     The particle on the left-hand side belongs to $\G\Short$, consisting of predominantly short cracks, while the particle on the right-hand side belongs to $\G\Long$, exhibiting long cracks.}
     \label{fig:ExperimentalLongAndShort}
 \end{figure}

By visual inspection of  segmented particles, it becomes appearant that the crack networks of some particles consist predominantly of short and others of long cracks, see Figures \ref{fig:SEMCrossSection} and~\ref{fig:ExperimentalLongAndShort}. Motivated by these  morphological differences, the   set $\G$ is subdivided into two disjoint subsets, $\G_\text{short}$ and $\G_\text{long}$.  To decide for a given particle $P_\text{ex}$ if it belongs to $\G\Short$ or $\G\Long$, a skeletonization algorithm \cite{LEE1994} was applied to the crack phase  of $P_\text{ex}$, where each connected component  of the crack phase $\crack$ is represented by its center line, which we refer to as a skeleton segment. The family  of all skeleton segments of a particle $P_\text{ex}$ is called skeleton  and denoted by~$\skel(P_\text{ex})$.

If the longest crack skeleton segment of a particle $P_\text{ex}$ is shorter than or equal to $t\cdot  \aed(P_{\text{ex}})$ for some threshold $t>0$,  then $P_\text{ex}$ is assigned to  $\G\Short$, otherwise $P_\text{ex}$ is assigned to $\G\Long$.  
Note that the area-equivalent diameter $\aed(P)$ of particle $P_\text{ex}$  is given by 
\begin{equation*}
    \aed(P_\text{ex})=\sqrt{\frac{4\,\nu_2(\solid\cup\, \crack)}{\pi}}\,,
\end{equation*}
where $\nu_2(A)$ denotes the 2-dimensional Lebesgue measure, i.e., the area of a set $A\subset\R^2$. Thus, formally, the sets $\G_\text{short}$ and $\G_\text{long}$
 can be written as 
\begin{align*}
    \G_\text{short} &= \{ P_\text{ex} \in \G \colon \max_{S\in \skel(P_\text{ex})} \mathcal{H}_1(S) \leq t \cdot \aed(P_\text{ex}) \} \quad \text{and}  \quad \G_\text{long}=\G \setminus \G_\text{short}
\end{align*} 
where $\mathcal{H}_1(S)$  denotes the 1-dimensional Hausdorff measure of a skeleton segment $S\in\skel(P_\text{ex})$, which corresponds to the length of $S$ \footnote{The length of a skeleton segment was approximated by the number of pixels multiplied with the resolution of $\rho=\SI{14.29}{\nano \meter}$.}.  
It turned out that $t=0.55$ is a reasonable choice, which splits $\G$ into two  subsets $\G_\text{short}$ and $\G_\text{long}$, each containing  particles from the entire range of observed particle sizes, where $\G\Short$ comprises 423 particles  and $\G\Long$ consists of 83 particles. For larger values of $t$ the statistical representativeness of  $\G\Long$ diminishes, whereas for smaller values of $t$ we observed that the resulting set $\G\Long$ was comprised of particles with relatively small cracks---which would have made the decomposition of particles into  $\G\Short$ and $\G\Long$ redundant.

\section{Stochastic 3D model for cracked NMC particles} \label{sec:Model}
In this paper, a stochastic 3D model is proposed, which generates  cracks in (simulated) pristine MNC particles hierarchically on different length scales. Two different kinds of data are used as model inputs.   First, to generate pristine NMC particles in 3D, exhibiting a polycrystalline inner structure, we draw samples from the stochastic particle model introduced in \cite{Furat2021}. Then, we use 2D SEM image data  to stereologically calibrate a stochastic 3D model for adding cracks, where we assume that cracks propagate along the polycrystalline grain boundaries through the particles without having a preferred direction. 
It is important to emphasize that the proposed stochastic crack model generates virtual, but realistic  cracked NMC particles in 3D, even though it is calibrated using only 2D image data.

In Section \ref{ssec:PristineModel}, the main features of the stochastic 3D model proposed in \cite{Furat2021} are summarized, which is used to generate the virtual, pristine NMC particles. 
To efficiently represent the neighborhood relationships of individual grains of a particle, in Section \ref{ssec:GraphBasedTessellation} a graph-based data structure is introduced  by means of Laguerre tessellations. 
Subsequently, in Section \ref{ssec:SingleCrackModel},  a stochastic model is presented, which incorporates  single cracks into the (previously simulated) pristine NMC particles, utilizing the  graph-based representation via tessellations stated in  Section \ref{ssec:GraphBasedTessellation}.  
Then, in Section  \ref{ssec:CrackEnsembleModel}, it is shown how  the single-crack model can be applied multiple times  to generate a random crack network consisting of several cracks within a  given particle. 
Finally, in Section \ref{ssec:CrackNetworkModel}, an extended stochastic crack network model is presented, which  is deployed for modeling the entire crack phase of NMC particles in 3D.

\subsection{Stochastic 3D model for pristine polycristalline NMC particles}\label{ssec:PristineModel}
In \cite{Furat2021} a spatial stochastic  model for the 3D morphology of pristine polycristalline NMC particles has been developed and calibrated by means of tomographic image data. More precisely, nano-CT data depicting the outer shell of  NMC particles has been leveraged to calibrate a  random field model on the three-dimensional sphere, whose realizations are virtual outer shells of NMC particles that are statistically similar to those observed in the nano-CT data. 
Furthermore, a random Laguerre tessellation model for the inner grain architecture of  NMC particles (which lives on a smaller length scale) has been calibrated using 3D EBSD data.

Note that a Laguerre tessellation in $\R^3$ is a subdivision of the three-dimensional Euclidean space (or some sampling window within $\R^3$) that is given by some marked point pattern $\{(s_n, r_n), n\in\N\}$, where $s_n\in\R^3$ is called a seed or generator point, and $r_n\in\R$ is an (additive) weight, for each $n\in\N=\{1,2,\ldots\}$, see \cite{Okabe00,chiu.2013}. The interior of the  grain\
generated by the $n$-th marked seed point $(s_n,r_n)$ of a Laguerre tessellation
 is defined as set of points $x\in\R^3$, which are  closer to $s_n$ than to all other seed points $s_k$, $k\neq n$, with respect to the \enquote{distance function} $d:\R^3\times\R^4\to\R$ given by $d(x,(s,r))=|x-s|-r$ for all $x,s\in\R^3$ and $r\in\R$, where $|  \cdot  |$ denotes the Euclidean norm in $\R^3$.\footnote{In the mathematical literature, the grains of a Laguerre tessellation are often called \enquote{cells}.  However,  for  modeling the polycrystalline materials considered in the present paper,  the wording \enquote{grain} is used.} Thus, formally, the grain $g_n\subset\R^3$  
generated by the $n$-th marked seed point $(s_n,r_n)$ 
 is given by
	\begin{align}
		g_n=\Bigl\{x\in\R^3 \  : \  d(x,(s_n,r_n))\leq d(x,(s_k,r_k)) \quad \text{ for all } k\neq n\Bigr\}.
\label{eq:LaguerreTessellation}
	\end{align}
To compute grains $g_n$ for a given set of marked seed points we use the GeoStoch library \cite{mayer2004unified}.

Both stochastic models mentioned above,  i.e., the random field model for the outer shell and the random tessellation model for the inner grain architecture,  have been combined in \cite{Furat2021}, to derive a multi-scale 3D model for pristine NMC particles with full inner grain architecture. Thus, in the first modeling step of the present paper, we will draw realizations  from the multi-scale 3D model of \cite{Furat2021} for the generation of virtual pristine NMC particles, to which cracks will be added in the subsequent modeling steps.
Using an analogous notation like that considered in Eq.~\eqref{for.con.rep},
the simulated pristine NMC particles  will  be denoted by 
 \mbox{$P_\text{pr}=(\solid^\prUp,\emptyset)$}, where
\begin{align*}
    \solid^\prUp=\bigcup_{n\in I} g_n \subset \R^3
\end{align*}
for some index set $I\subset\N$.  
The stochastic crack model introduced later on (in Sections~\ref{ssec:SingleCrackModel} to~\ref{ssec:CrackNetworkModel}) assigns facets, i.e. planar grain boundary segments,  of the pristine particle $P_\text{pr}$  with crack widths to introduce a crack network. 
To do so, we first derive an alternative graph representation of the Laguerre tessellation $\{g_n,n\in I\}$ which describes the  grain architecture of $P_\text{pr}$.

\subsection{Graph representation of pristine  grain architectures}\label{ssec:GraphBasedTessellation}
In the literature, a Laguerre tessellation in $\R^3$ is usually given by a collection of grains $g_n\subset\R^3$ as defined in Eq.~\eqref{eq:LaguerreTessellation}. However,
alternatively,  such a  tessellation 
can be represented as a collection of planar facets given by 
\begin{equation*}\label{def.pla.fac}
g_n\cap g_k=\{x\in\R^3 \  : \  d(x,(s_n,r_n))= d(x,(s_k,r_k))\}
\end{equation*}
for  $n,k\in\N$ with $n\not= k$ and $\mathcal{H}_2(g_n\cap g_k)>0$, where  $\mathcal{H}_2(g_n\cap g_k)$ is the 2-dimensional Hausdorff measure of $g_n\cap g_k\subset\R^3$, which corresponds to the area of $g_n\cap g_k$.
Thus,
the sets $g_n\cap g_k$ are convex  plane segments 
being the intersection of  neighboring grains,
the union of which is equal to the union of the boundaries 
$\partial g_n$
of the  convex polyhedra $g_n$ considered in Eq.~\eqref{eq:LaguerreTessellation}.

Furthermore, to describe the neighborhood structure of the facets, we consider the so-called neighboring facet graph, denoted by $G=(F,E)$. The set  $F$ of its vertices is the collection of planar  facets of the Laguerre tessellation, and $E$ is its set of edges, where two facets $f,f^\prime\in F$ are connected by an edge $e\in E$ if they are adjacent, which means that  $f\cap f^\prime$ is a line segment with positive length, i.e., $\mathcal{H}_1(f\cap f^\prime)>0$, see Figures \ref{fig:GraphWorkflow}a and~\ref{fig:GraphWorkflow}b.

\subsection{Single crack model}\label{ssec:SingleCrackModel}
\phantom{s}
In this section, we describe the stochastic model which will be used for the insertion of  single cracks into  virtual NMC particles,   whose polycristalline grain architecture  is given by a Laguerre tessellation within a certain (bounded) sampling window $W\subset\R^3$, as stated in Section~\ref{ssec:PristineModel}, and represented by the neighboring facet graph $G=(F,E)$  introduced in Section \ref{ssec:GraphBasedTessellation}.

\begin{figure}[h]
    \centering
    \includegraphics[width=0.55\textwidth]{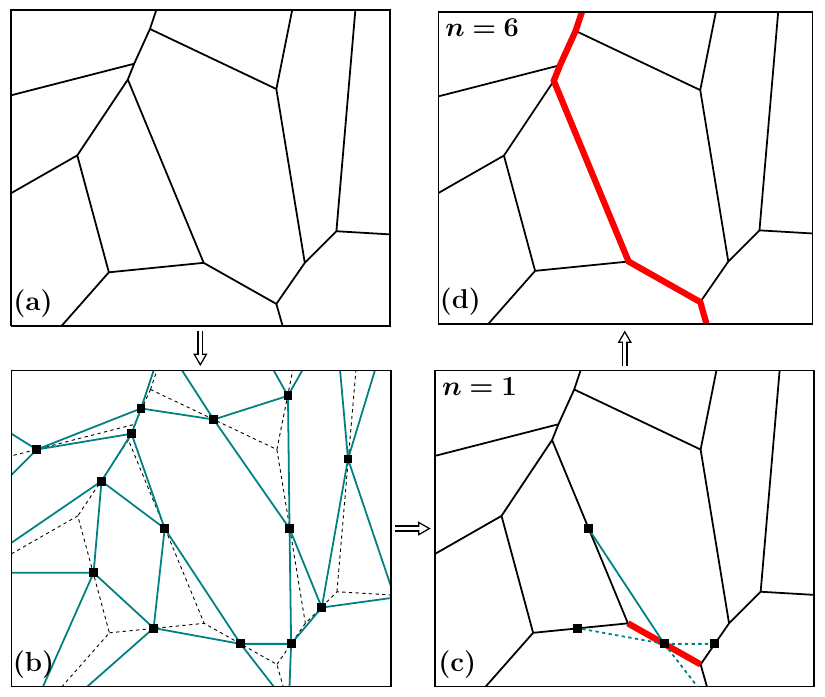}
    \caption{2D scheme of the workflow to generate an individual crack along grain boundaries. For a (Laguerre) tessellation within a bounded sampling window (a), the neighboring facet graph is determined, i.e., facets are considered to be vertices of the graph (black rectangles), which are  connected by  edges (blue) if the underlying facets are adjacent (b).  An initial facet (red) is  chosen at random and assigned to the set $C$ of crack facets (c). Iteratively, the $n$-th neighboring facet which is aligned \enquote{best} with the set  $C$ is assigned to it (d).}
    \label{fig:GraphWorkflow}
\end{figure}

Assuming that  cracks propagate along  grain boundaries \cite{LSMCM23}, we will model cracks as  collections of dilated adjacent facets.  With regard  to the graph-based representation of  tessellations stated in Section \ref{ssec:GraphBasedTessellation},  this means that  a subset $C\subset F$ will be chosen such that for each pair  $f,f^\prime\in C$ with $f\not= f^\prime$,  
there exists a sequence of adjacent facets $f_1,\ldots,f_{n}\in C$ such that $f_1=f$ and $f_n=f^\prime$. 
Note that this method can generate long contiguous cracks, spanning along adjacent facets, even with a relatively small number of cracked facets. This allows generating, if desired, particles with a relatively low quantity of cracked facets, but with relatively long contiguous cracks, which would be not possible with a stochastic approach not considering sequence of adjacent facets. Then, in a second step, each crack facet $f\in C$ will be morphologically dilated to a specific thickness, see Figure \ref{fig:GraphWorkflow}. In this approach, a facet between two grains is either pristine or fully cracked.

More precisely, to generate a set of dilated crack facets as described above, an algorithm is proposed consisting of the following steps:
\begin{itemize}
    \item[(i)] Initialize the set of crack facets,  putting $C=\emptyset$.
    \item[(ii)] Generate the number $n_\text{facets}\in\N$ of crack facets, drawing a realization $ \hat{n}_\text{facets}>0$
 from a Weibull distributed random variable $N_\text{facets}$ with some scale parameter $\lambda\WeiInd>0$ and shape parameter $k\WeiInd>0$, 
and putting $n_\text{facets}=\text{round}(\hat{n}_\text{facets})$, where 
\begin{align}
    \text{round}(\hat{n}_\text{facets})=
    \begin{cases}
        \lfloor \hat{n}_\text{facets}\rfloor \quad &\text{if} \qquad   \hat{n}_\text{facets} -\lfloor\hat{n}_\text{facets}\rfloor  < 0.5, \\
        \lfloor \hat{n}_\text{facets}\rfloor+1 &\text{else,}
    \end{cases}
    \label{eq:defRoundToInt}
\end{align}
which means rounding to the closest integer, with $\lfloor \hat{n}_\text{facets}\rfloor$ denoting the largest integer being smaller than  $\hat{n}_\text{facets}$.
    \item[(iii)] Choose an  initial facet $f\in F$ at random and assign it to the set of crack facets $C$.  
Furthermore, let $g:F\rightarrow\R^3$ denote a function, which maps a facet $f\in F$ onto its  normal vector $v=(v_1,v_2,v_3)$ with length  1 and  $v_1\geq0$. 
    \item[(iv)] Compute the average normal vector 
$v_C=\sum_{f\in C} g(f)\bigl|\sum_{f\in C} g(f)\bigr|^{-1},
$ 
to control the alignment of the next facet, to be assigned to $C$.  
    \item[(v)] Determine the set 
$A=\{f \in F \setminus C \colon f\cap f^\prime \in E \text{ for some } f^\prime \in C\}\subset F\setminus C$, containing the facets that are adjacent to $C$, but not contained in $C$.
    \item[(vi)] Add  the facet $f \in A$ given by 
 $ f=\argmax_{f\in A} \ |\langle g(f),v_C \rangle|
    \label{eq:NextFacetCritereon}$
 to $C$, for which the normal $g(f)$ has the best directional alignment with the average normal vector $v_c$ computed in step (iv), 
where $\langle\cdot,\cdot\rangle$ denotes the dot product.
    \item[(vii)] Repeat steps (iv) to (vi) until $\#C=n_\text{facets}$, where $\#$ denotes  cardinality. 
    \item[(viii)] Draw a realization $\delta>0$ from  a gamma distributed random variable $\Delta$ with some shape parameter $k_\Gamma>0$ and scale parameter $\theta_\Gamma>0$.
    \item[(ix)]  Dilate each crack facet $f\in C$   using the  structuring element   $B_{\delta}=\{x\in\R^3:|x|\le\delta/2\}$, and determine the set $\bigcup_{f\in C} (f\oplus B_{\delta})$, where $\oplus$ denotes Minkowski addition. 
Note that the set $\bigcup_{f\in C} (f\oplus B_{\delta})$ represents a crack where each facet $f\in C$ is dilated with the same thickness $\delta$. 
\end{itemize}

In summary, the stochastic model  for  single cracks described above is characterized by the 4-dimensional parameter vector $\thetaBar=(\lambda\WeiInd, k\WeiInd, k_\Gamma, \theta_\Gamma)\in\R_+^4$, where $\lambda\WeiInd$ and $ k\WeiInd$ control the length of cracks, whereas $k_\Gamma$ and $ \theta_\Gamma$ affect their thickness.  

Recall that in Section~\ref{ssec:PristineModel} we introduced the notation $P_\text{pr}=(\solid^\prUp,\emptyset)$ for simulated pristine NMC particles. Analogously, for a given pristine particle $P_\text{pr}=(\solid^\prUp,\emptyset)$,
a particle with a single crack will be denoted by $P_{\thetaBar}=(\solid^{(\thetaBar)},\crack^{(\thetaBar)})$, where  $\solid^{(\thetaBar)}\cup \, \crack^{(\thetaBar)}=\solid^\prUp$, with $\solid^{(\thetaBar)},\crack^{(\thetaBar)}\subset\R^3$
being the solid and crack phase of $P_{\thetaBar}$, respectively.
 More precisely, it holds that  
\begin{align*}
 \crack^{(\thetaBar)}&=\Bigl\{
 x\in\solid^\prUp\colon  \dist(x,f)\leq \frac{\delta}{2} \text{ for some } f \in C 
 \Bigr\} 
\end{align*}
and 
$ \solid^{(\thetaBar)}=
\solid^\prUp\setminus\crack^{(\thetaBar)}$,
where 
$\dist(x,f)=\min\{|x-y|: y\in f\}$ denotes the Euclidean distance from $x\in\R^3$  to the set $f\in C$.

\subsection{Crack network model}\label{ssec:CrackEnsembleModel}
Typically, the crack phase of particles observed in experimental 2D SEM data consists of more than one crack and forms complex crack networks, see Figure~\ref{fig:SEMCrossSection}.

Thus, to model the crack phase of  particles consisting of multiple cracks, we draw a realization $\nCracks\in \N\cup\{0\}$ from a Poisson distributed random variable $\NCracks$ with some parameter $\lambda\PoiInd>0$. 
Furthermore, let 
$P_{\thetaBar,1},
\ldots,
P_{\thetaBar,\nCracks}$ with $P_{\thetaBar,i}=(\solid^{(\thetaBar,i)},\crack^{(\thetaBar,i)})$ for $i=1,\ldots,\nCracks$ denote independent realizations of the single crack model introduced in Section~\ref{ssec:SingleCrackModel}, applied to one and the same pristine particle $P_\text{pr}=(\solid^\prUp,\emptyset)$. Overlaying these realizations results in a realization of the crack network model $P_{\theta_2}=(\solid^{(\theta_2)},\crack^{(\theta_2)})$ with parameter vector 
\begin{equation}\label{def.the.two}
\theta_2=(\thetaBar,\lambda\PoiInd)= (\lambda\WeiInd, k\WeiInd, k_\Gamma, \theta_\Gamma,\lambda\PoiInd) \in\R^5_+ ,
\end{equation} 
where
$\crack^{(\theta_2)}=\bigcup_{i=1}^{\nCracks} \crack^{(\thetaBar,i)}$ 
    and $\solid^{(\theta_2)}=\solid^\prUp\setminus\crack^{(\theta_2)}$. 

By visual inspection of the SEM data, see Figure \ref{fig:SEMCrossSection}, it is obvious that the distributions of  the random number $\NCracks$ and size $N_\text{facets}$ of individual cracks should depend on  the  size  of the underlying pristine particle $P_\text{pr}$, i.e.,  small particles tend to have less and shorter cracks, whereas large particles exhibit more and longer cracks. Therefore, we assume that the scale parameters $\lambda\PoiInd,\lambda\WeiInd>0$ considered in Eq.~\eqref{def.the.two} are given by
\begin{equation}\label{rep.sca.par}
    \lambda\PoiInd= \lambda\PoiInd(c\PoiInd,c\DimInd)= c\PoiInd \nubig{\solid^\prUp}^{c\DimInd}, \qquad \lambda\WeiInd= \lambda\WeiInd(c\WeiInd,c\DimInd)= c\WeiInd \nubig{\solid^\prUp}^{1-c\DimInd}
\end{equation}
for some constants $c\PoiInd,c\WeiInd>0$ and $c\DimInd\in[0,1]$, where $\nu_3$ denotes the 3-dimensional Lebesgue measure, i.e., $\nubig{\solid^\prUp}$ is the volume of $P_\text{pr}$. This implies that the porosity
\begin{equation*}
p=\frac{\mathbb{E}\nubig{\crack^{(\theta_2)}}}{\nubig{\solid^\prUp}} ,
\end{equation*}
of the crack network model $P_{\theta_2}$ does not (or only slightly) depend on  the  volume $\nubig{\solid^\prUp}$  of the underlying pristine particle $P_\text{pr}$, which can be shown as follows.  
Since the random variables $\NCracks, N_\text{facets}, \Delta$ are assumed to be independent, it holds that\footnote{Note that this approximation does not take the overlap of cracked facets into consideration.}
\begin{align}
p=\frac{\mathbb{E}\nubig{\crack^{(\theta_2)}}}{\nubig{\solid^\prUp}} 
    &\approx \frac{\alpha\mathbb{E}\NCracks\mathbb{E}\nFacets \mathbb{E}\Delta  }
    {\nubig{\solid^\prUp}} 
    =\frac{\alpha\lambda\PoiInd\lambda\WeiInd \, \gamma_{k\WeiInd} \mathbb{E}\Delta  }{\nubig{\solid^\prUp}},
    \label{eq:Porosity.neu}
\end{align}
where $\alpha>0$ is the mean area of planar facets of the Laguerre tessellation describing the grain architecture of $P_\text{pr}$ and $\gamma_{k\WeiInd}=\Gamma(1+\frac{1}{k\WeiInd})$ with the Gamma function $\Gamma\colon(0,\infty)\mapsto\R_+$ given by $\Gamma(x)=\int_0^\infty t^{z-1}e^{-t}dt$. 
Thus, inserting Eq.~\eqref{rep.sca.par} into Eq.~\eqref{eq:Porosity.neu}, we get that
\begin{align*}
    p 
    &\approx 
    \frac{\alpha \,c\PoiInd\nubig{\solid^\prUp}^{c\DimInd}\, c\WeiInd\nubig{\solid^\prUp}^{1-c\DimInd}\, \gamma_{k\WeiInd}\mathbb{E}\Delta  }
    {\nubig{\solid^\prUp}}\\
    &= \nubig{\solid^\prUp}\ \frac{\alpha\, c\PoiInd\, c\WeiInd\, \gamma_{k\WeiInd}\mathbb{E}\Delta  }
    {\nubig{\solid^\prUp}}
    = \, \alpha\, c\PoiInd\, c\WeiInd\, \gamma_{k\WeiInd}\mathbb{E}\Delta,
\end{align*}
i.e., the porosity $p$ of the crack network model $P_{\theta_2}$ does not (or only slightly) depend on  the  volume $\nubig{\solid^\prUp}$  of the underlying pristine particle $P_\text{pr}$.

Finally, we remark  that from now on, utilizing the representation of the scale parameters  $\lambda\PoiInd$ and $\lambda\WeiInd$ introduced in Eq.~\eqref{rep.sca.par}, the following modified form of the parameter vector ${\theta_2}$ of $P_{\theta_2}$ given in Eq.~\eqref{def.the.two} will be used:
\begin{equation}\label{wri.the.two}
\theta_2=(c\WeiInd,k\WeiInd,k_\Gamma,\theta_\Gamma,c\PoiInd,c\DimInd)\in\R^5_+\times[0,1].
\end{equation}

\subsection{Extended crack network model}\label{ssec:CrackNetworkModel}
Recall Section \ref{ssec:Subdivison}, where the set of experimentally measured 2D SEM images $\G$ was split into two classes, $\G\Short$ and $\G\Long$, containing particle cross sections showing either predominantly short or long cracks. Nevertheless, each crack network exhibited on these cross sections, still consists of both, (relatively) short as well as (relatively) long cracks, see Figure \ref{fig:ExperimentalLongAndShort}. 

This is the reason why the crack network model that was introduced in Section~\ref{ssec:CrackEnsembleModel} turns out to be insufficiently flexible.  Therefore, we extend this model by realizing it twice on the same pristine particle $P_\text{pr}=(\solid^\prUp,\emptyset)$, with two different parameter vectors 
\begin{align*}
\theta_2\ShortUp&=(c\WeiInd\ShortUp,k\WeiInd\ShortUp,k_\Gamma\ShortUp,\theta_\Gamma\ShortUp,c\PoiInd\ShortUp,c\DimInd\ShortUp) \qquad \text{and}\qquad
\theta_2\LongUp=(c\WeiInd\LongUp,k\WeiInd\LongUp,k_\Gamma\LongUp,\theta_\Gamma\LongUp,c\PoiInd\LongUp,c\DimInd\LongUp).
\end{align*}
In this way we obtain two independently cracked particles 
\begin{align*}
P_{\theta\ShortUp_2}=(\solid^{\theta\ShortUp_2},\crack^{\theta\ShortUp_2})\qquad \text{and}\qquad P_{\theta\LongUp_2}=(\solid^{\theta\LongUp_2},\crack^{\theta\LongUp_2}),
\end{align*}
which are used to get the extended crack network model $P_\theta=(\solid^\thetaUp,\crack^\thetaUp)$ with $\theta=(\theta\ShortUp_2,\theta\LongUp_2)$, exhibiting a sufficiently large variety of short and long cracks, where
\begin{equation}\label{def.xii.the}
 \crack^\thetaUp= \crack^{(\theta_2\ShortUp)}\cup\crack^{(\theta_2\LongUp)}
 \qquad \text{and}\qquad
 \solid^\thetaUp= \solid^\prUp\setminus\crack^\thetaUp.
\end{equation}

By visual inspection of the segmented SEM data, see Figure \ref{fig:ExperimentalLongAndShort}, it seems clear that short and long cracks  exhibit  similar thicknesses. This observation motivates a reduction of model parameters by setting 
$k_\Gamma=k_\Gamma\ShortUp=k_\Gamma\LongUp$ and 
$\theta_\Gamma=\theta_\Gamma\ShortUp=\theta_\Gamma\LongUp$.
Furthermore, we assume that the influence of the volume $\nubig{\solid^\prUp}$  of the underlying pristine particle $P_\text{pr}$ on the distributions of the  number and size of cracks is the same for short and long cracks, i.e., we assume that  $c\DimInd\ShortUp=c\DimInd\LongUp=c\DimInd$. Thus, the number of model parameters is reduced from 12 to 9, leading to the parameter vector
\begin{equation}\label{eq:finalParamVector}
    \theta=(
    c\WeiInd\ShortUp,k\WeiInd\ShortUp,c\PoiInd\ShortUp,
    c\WeiInd\LongUp,k\WeiInd\LongUp,c\PoiInd\LongUp,
    k_\Gamma,\theta_\Gamma,c\DimInd)
    \in\R_+^8\times[0,1]    
\end{equation}
of the extended crack network model, where $c\WeiInd^{(i)},k\WeiInd^{(i)}$ control the  length, $c\PoiInd^{(i)}$ the number and $k_\Gamma,\theta_\Gamma$ the thickness of cracks for  $i\in\{1,2\}$, whereas $c\DimInd$ controls the influence 
 of the volume $\nubig{\solid^\prUp}$  of  $P_\text{pr}$ on the distributions of the  number and size of cracks. 
	
\section{Model calibration}\label{ssec:Fitting}

The calibration of the extended crack network model proposed in Section~\ref{ssec:CrackNetworkModel}  is organized as follows.
First, in Section~\ref{ssec:MinProb}, we formulate a minimization problem to determine optimal values of the parameter vector $\theta$ given in Eq.~\eqref{eq:finalParamVector}. For this, in Section~\ref{sssec:descriptors},  three different geometric descriptors of image data are introduced. These geometric descriptors are used in Section~\ref{sssec:lossfunction} to define a loss function, which measures the discrepancy between experimentally imaged particle cross sections and those drawn from the  crack network model.  Finally, in Section \ref{sssec:Training}, a numerical method is described for solving the minimization problem stated in Section~\ref{ssec:MinProb}.
 
\subsection{Minimization problem}\label{ssec:MinProb}
\begin{figure}[h]
    \centering
    \includegraphics[width=.7\textwidth]{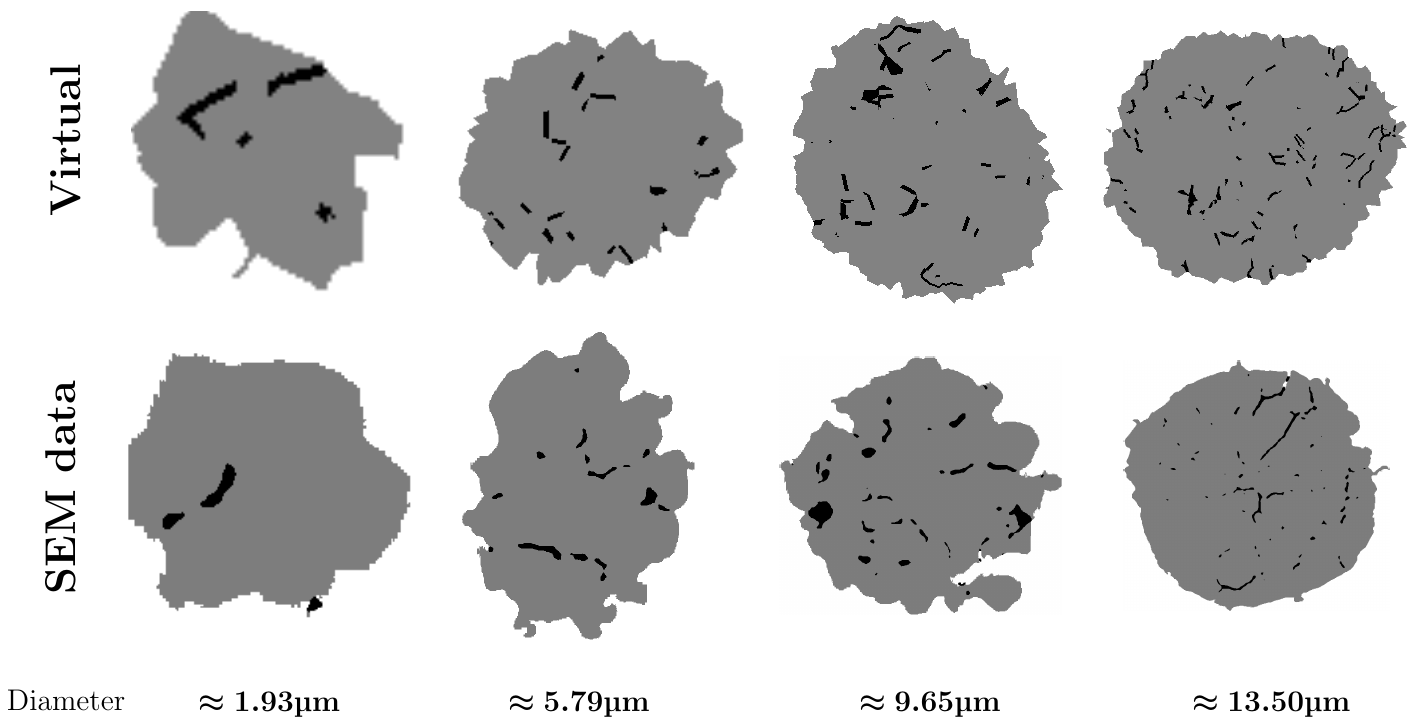}
    \caption{Particle cross sections across various size classes, drawn from the extended crack network model calibrated to $\G\Short$ (upper row) and corresponding representatives of $\G\Short$ (lower row).
    The cross sections were scaled to the same size, while their actual sizes are indicated by their  area-equivalent diameters.
    }
    \label{fig:VisualShort}
\end{figure}

\begin{figure}[h]
    \centering
    \includegraphics[width=.7\textwidth]{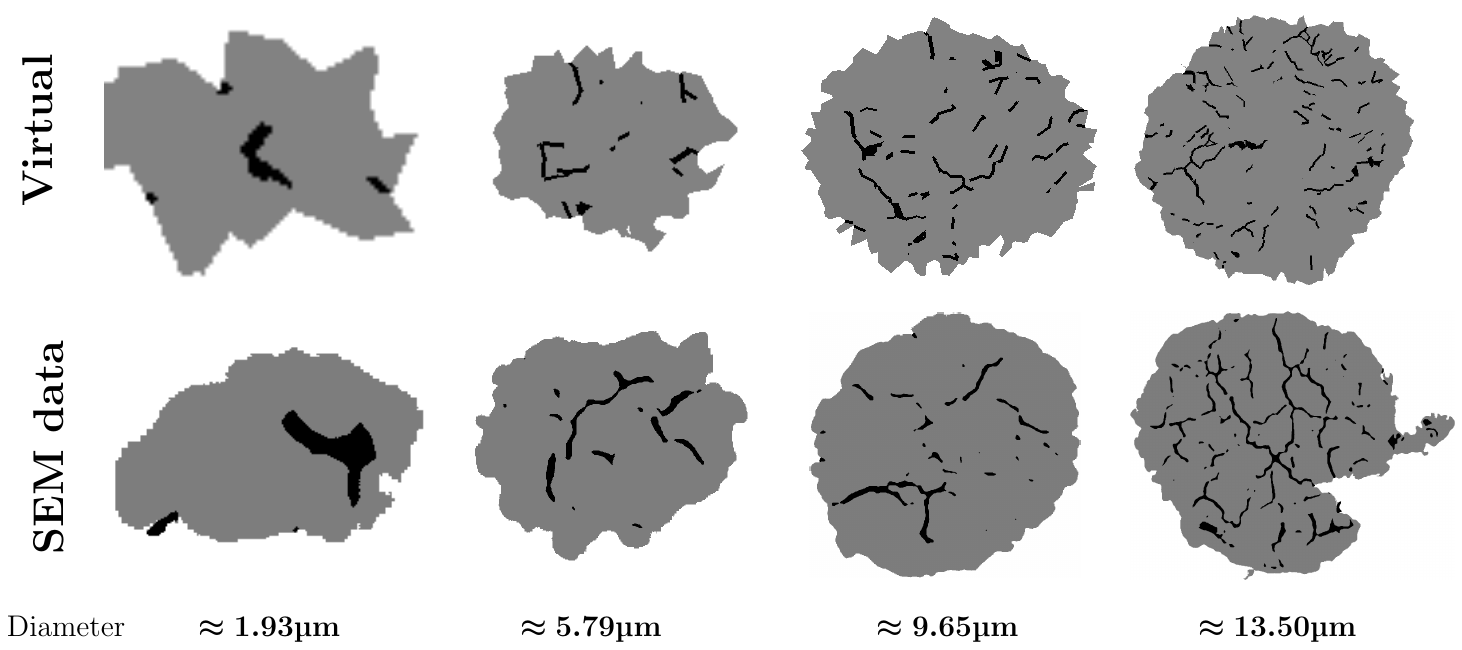}
    \caption{Particle cross sections across various size classes, drawn from the extended crack network model calibrated to $\G\Long$ (upper row) and corresponding representatives of $\G\Long$ (lower row).
    The cross sections were scaled to the same size, while their actual sizes are indicated by their  area-equivalent diameters.
    }
    \label{fig:VisualLong}
\end{figure}

The extended crack network model parameters introduced in Section \ref{ssec:CrackNetworkModel} are separately fitted to both partitions,  $\G\Short$ and $\G\Long$, of the experimental data set $\G$ considered in Section~\ref{ssec:Subdivison}. Thus, the optimization of the parameter vector $\theta$ given in Eq.~\eqref{eq:finalParamVector} is performed twice, for $\G\Short$ and $\G\Long$, where the discrepancy between geometric descriptors of experimental image data and simulated image data drawn from the extended crack network model is minimized. Figures~\ref{fig:VisualShort} and~\ref{fig:VisualLong} illustrate cross section realizations of virtual particles drawn from the extended crack network model fitted to $\G\Short$ and $\G\Long$, respectively, alongside experimentally imaged cross sections.

Furthermore, it is important to note that the crack network morphology may significantly vary across different cross-section sizes, see Figure \ref{fig:SEMCrossSection}.  To avoid systematic errors arising from comparing experimental and simulated cross sections of different sizes, we introduce several cross-section size classes. Thus, experimental and simulated cross-sections are only compared if they are approximately of the same size. More specifically, a simulated particle cross-section is compared to the average of all experimental cross-sections in the same size class.

For the sake of simplicity, 
we will use the following abbreviating notation, writing $\G$  instead of $\G\Short$ and $\G\Long$. Furthermore, for each $d>0$,  let $\restr{\G}{d}$ be the restriction  of $\G$ to all particle cross sections  $P_\text{ex}$ whose area-equivalent diameter $ \aed(P_\text{ex})$ belongs to the interval $ B_\ell(d)=[j \ell, (j+1)\ell) $ with given length $\ell>0$, where the integer $j\in\N\cup\{0\}$ is chosen such that  $d\in [j \ell, (j+1)  \ell)$. It turned out that an interval length of $\ell\approx\SI{1.29}{\micro\meter}$ 
 balances  a reasonable number of experimental cross sections in each bin and, simultaneously,  preserves a sufficiently fine subdivision of the entire dataset $\G$,  where this subdivision results into 11 size intervals $[0,\ell),\ldots[10\ell,11\ell)$, with $11\ell\approx \SI{14.1}{\micro\meter}$.\footnote{The interval length of $\ell\approx\SI{1.29}{\micro\meter}$ corresponds to approximately 90 pixels of the experimental data.}  However,
 since the stochastic 3D model for pristine NMC particles,  described in Section~\ref{ssec:PristineModel},
 has been calibrated to 3D nano-CT data \cite{Furat2021}, it happens that for some randomly oriented planes $E\subset\R^2$, the cross sections $P_\theta\cap E$ of    3D particles drawn from the extended crack network model $P_\theta$  are larger than the ones observed in the dataset $\G$, which were measured by the 2D SEM technique. Thus, if the area-equivalent diameter of $P_\theta\cap E$    is larger than $11\ell$, which is the upper bound of the largest size class $B_\ell(10)$ of the experimental data set $\G$ (for both cases $\G=\G\Short$ and $\G=\G\Long$),  then $P_\theta\cap E$   is not considered in the  minimization procedure.

 This leads to the minimization problem
\begin{align}
    \widehat{\theta}=\argmin_{\theta\in\R_+^8\times[0,1]}\quad \EX L\Bigl(P_{\theta}\cap E, \restr{\G}{\aed(P_{\theta}\cap E)}\Bigl),
    \label{eq:LossFunction}
\end{align}
where the expectation  in Eq.~\eqref{eq:LossFunction} extends over cross sections $P_\theta\cap E$ such that $\aed(P_{\theta}\cap E)\le 11\ell$, and $L(\cdot,\cdot)$ is some loss function, which measures the discrepancy between  the cross section $P_{\theta}\cap E$ of the extended crack network model $P_\theta$ and  particle cross sections belonging to the restriction $ \restr{\G}{\aed(P_{\theta}\cap E)}$ of $\G$.

\subsection{Geometric descriptors of 2D image data}\label{sssec:descriptors}
In this section, three different geometric descriptors of 2D image data are considered: the two-point coverage probability function, the crack-size distribution and the distance-to-background distribution. They will be determined on (measured and simulated)  particle cross sections, denoted by $P=(\solid,\crack)$, where $\solid,\crack\subset\R^2$. 
Furthermore, these descriptors will be employed in Section~\ref{sssec:lossfunction} to determine the loss function considered in Eq.~\eqref{eq:LossFunction}.

\vspace{3mm}
\paragraph{Two-point coverage probability} \label{sssec:tpp}
For each $h\in[0,h_{\max}]$, where $h_{\max} >0$ is some maximum distance,  the so-called  the two-point coverage probability, denoted by $\TPP_{\Xi}(h)$, is the probability that two randomly chosen points $x_1,x_2\in\solid\cup\crack$ of distance $h$ belong to the particle phase $\Xi\in\{\solid,\crack\}$. This probability will be estimated by  the number of pixel pairs $x_1,x_2\in\Xi\cap\Z^2$  separated by  distance $h$, divided by the total number of  pixel pairs $x_1,x_2\in(\solid\cup \crack)\cap\Z^2$ of distance $h$, i.e., 
\begin{align*}
    \TPP_{\Xi}(h)\approx \frac{\# \{ x_1,x_2\in\Xi\cap\Z^2\colon | x_1-x_2 |=h \} }
    {\# \{ x_1, x_2\in(\solid\cup\crack)\cap\Z^2\colon | x_1-x_2 |=h \}},
\end{align*}
where $|\,\cdot\,|$ denotes the Euclidean norm in $\R^2$, see e.g., \cite{chiu.2013} for further details. 

For the data considered in the present paper, the two-point coverage probabilities $\TPP_{\solid}(h)$ and $\TPP_{\crack}(h)$ are estimated for all possible distances $h\in[0,h_\text{max}]$ on the pixel grid, where $h_\text{max}\approx\SI{850}{\nano \meter}$, because  it turned out that the values obtained for $\TPP_{\solid}(h)$ and $ \TPP_{\crack}(h)$ are typically constant
  for $h>\SI{850}{\nano \meter}$. 
These probabilities are then interpolated utilizing cubic splines and evaluated for 30 equidistant values of  $h$, corresponding to a step size of approximately $\SI{28}{\nano\meter}$, which  leads to the  vectors of relative frequencies
\begin{align}\label{his.two.sol}
\TPP_\solidText(P)&=\bigl(\TPP_{\solid}(h_0),\TPP_{\solid}(h_1), \ldots,\TPP_{\solid}(h_{29})\bigr)\in[0,1]^{30} 
\end{align}
and 
\begin{align}\label{his.two.cra}
\TPP_\crackText(P)&=\bigl(\TPP_{\crack,}(h_0),\TPP_{\crack,}(h_1),\ldots,\TPP_{\crack}(h_{29})\bigr)\in[0,1]^{30},
\end{align}
where $h_i=i\,h_{\max}/29$ for $i=0,1,\ldots,29$.

\vspace{3mm}
\paragraph{Crack-size distribution} \label{sssec:crackSizeDistribution}
The probability distribution of the size of a randomly chosen crack  within a particle cross section $P=(\solid,\crack)$ will also be incorporated into the loss function introduced  in Eq.~\eqref{eq:LossFunction}. Formally, a crack is considered to be a connected component of the   crack phase $\crack$, where the crack size will be represented by the area-equivalent diameter of the crack.

Recall that $\aed(\xi)$ denotes the area-equivalent diameter of a set $\xi\subset\R^2$. Furthermore, let $\xi_1,\ldots,\xi_n\subset\crack$  denote the connected components of the crack phase $\crack$.  
The probability density of the random crack size will then be estimated by a histogram  with some bin width $w>0$,  which is given by  the relative frequencies
\begin{align*}
    \crackSize(k)=\frac{\#\bigl\{ \xi\in\{\xi_1,\ldots,\xi_n\} \colon \aed(\xi) \in [kw,(k+1)w) \bigr\}}{n}
\end{align*}
for $k=0,1,\ldots, 49$, where we put $w=\SI{50}{\nano\meter}$. Altogether, this leads to the vector of relative frequencies 
$   \crackSize(P)=\bigl(\crackSize(0),\crackSize(1),\ldots,\crackSize(49)\bigr)\in[0,1]^{50}.$

\vspace{3mm}
\paragraph{Distance-to-background distribution} \label{sssec:distanceToBackgroundDistribution}
Consider a randomly chosen point $X\in\crack$ within the crack phase $\crack$ of a particle cross section $P=(\solid,\crack)$, and the random (minimum) distance $D$ from $X$ to the background $\R^2\setminus (\solid\cup\crack)$ surrounding $P$, i.e., 
\begin{equation*}
D=\min\{|X-y|:y\in \R^2\setminus (\solid\cup\crack)\}. 
\end{equation*}
 The probability distribution of the random variable $D$  will  be taken into account as a third component  in the loss function introduced  in Eq.~\eqref{eq:LossFunction}.  Like for the crack sizes considered above, the probability density of $D$ will be estimated by a histogram with some bin width $w>0$, which is specified by the relative frequencies
\begin{align*}
    \distToBack(k)=\frac{\#\{ x\in\crack\cap\Z^2\colon 
    \min\{|x-y|:y\in \R^2\setminus (\solid\cup\crack)\}
\in[kw,(k+1)w) \}}{\#\crack\cap\Z^2}
\end{align*}
for $k=0,1,\ldots,119$, where we put $w=\SI{50}{\nano \meter}$.
In summary, we obtain the vector of relative frequencies
$\distToBack(P)=\bigl(\distToBack(0),\distToBack(1),\ldots,\distToBack(119)\bigr)\in[0,1]^{120}.$

\subsection{Loss function}\label{sssec:lossfunction}
We now specify the loss function $L(\cdot\,,\cdot)$ considered in Eq.~\eqref{eq:LossFunction}, utilizing the geometric particle descriptors stated in Section~\ref{sssec:descriptors}. Recall that the purpose of the loss function is to measure the discrepancy between experimentally imaged particle cross sections and those drawn from the  extended crack network model stated in Section~\ref{ssec:CrackNetworkModel}. In particular, the loss function will be utilized in Section \ref{sssec:Training} to solve the minimization problem introduced in Eq.~\eqref{eq:LossFunction}.

Let $\G$  denote some set of experimentally imaged particle cross sections, e.g. $\G=\G\Short$.
Furthermore, let  
\begin{align*}
   \avTPP_{\solidText}(\G)=\frac{1}{\#\G}\sum_{P=(\solid,\crack)\in\G} \tpp_\solidText(P).
\end{align*}
be the componentwise average  of the vector of relative frequencies given in Eq.~\eqref{his.two.sol}.
The averages $\avTPP_\crackText(\G), \avcrackSize(\G)$ and  $\avdistToBack(\G)$ for the two-point coverage probability of the crack phase, crack-size distribution and distance-to-background distribution are defined analogously. The loss function considered in Eq.~\eqref{eq:LossFunction} is then given by
\begin{align*}
		L\bigl(P_\theta\cap E,\restr{\G}{\aed(P_\theta)}\bigr)
		&=\mae\Bigl(\TPP_{\crackText}(P_\theta\cap E),\avTPP_{\crackText}\Bigl(\restr{\G}{\aed(P_\theta\cap E)}\Bigr)\Bigr)\\
		&\ +\mae\Bigl(\TPP_{\solidText}(P_\theta\cap E),\avTPP_{\solidText}\Bigl(\restr{\G}{\aed(P_\theta\cap E)}\Bigr)\Bigr)\\
		&\ +\mae\Bigl(\crackSize(P_\theta\cap E),\avcrackSize\Bigl(\restr{\G}{\aed(P_\theta\cap E)}\Bigr)\Bigr)\\
		&\ +\mae\Bigl(\distToBack(P_\theta\cap E),\avdistToBack\Bigl(\restr{\G}{\aed(P_\theta\cap E)}\Bigr)\Bigr),
\end{align*}
where $\mae(\cdot,\cdot)$ is an error function which quantifies the discrepancy between a random cross section $P_{\theta} \cap E$  
of the extended crack network model $P_{\theta}$, and the  set $\restr{\G}{\aed(P_\theta\cap E)}$ of experimentally imaged cross sections. More precisely, we consider the  truncated mean absolute error  
\begin{align}
    \mae(x,y)=\frac{1}{n_+}\sum_{i=1}^{n_+}|x_i-y_i|
    \label{eq:modiefiedMAE}
\end{align}
for $x=(x_1,\ldots,x_n),y=(y_1,\ldots,y_n)\in\R^n$,
where $n_+\le n$ is  the smallest integer $j\in\{1,\ldots,n\}$ such that  $x_i=y_i=0$ for all $i\in\{j+1,\ldots,n\}$.

Note that truncating the sum in Eq.~\eqref{eq:modiefiedMAE} at $n_+\le n$ is motivated by the fact that the components of the vectors of relative frequencies considered in Section~\ref{sssec:descriptors}  are equal to zero from a certain index. For the two-point coverage probabilities $\tpp_\text{solid}(\cdot)$ and $\tpp_\text{crack}(\cdot)$ occurring in Eqs.~\eqref{his.two.sol} and \eqref{his.two.cra}, respectively, this happens when the  size of the particle cross section is smaller than $h_\text{max}\approx\SI{850}{\nano \meter}$. Furthermore, for $\crackSize(\cdot)$ and $\distToBack(\cdot)$, some cross sections may contain only features smaller than a certain threshold. Truncating the sum in Eq.~\eqref{eq:modiefiedMAE} ensures that the sum of absolute values of the right-hand side of Eq.~\eqref{eq:modiefiedMAE} is  normalized with the actual number of non-zero components of both vectors $x,y\in\R^n$. Thus, this approach prevents that the error considered in Eq.~\eqref{eq:modiefiedMAE} is not appropriately weighted, which could occur if many components of $x,y\in\R^n$ are equal to  zero.

\subsection{Numerical solution of the minimization problem}\label{sssec:Training}
%\paragraph{\textbf{Training procedure}} \label{sssec:Training}
For solving the minimization problem stated in Eq.~\eqref{eq:LossFunction},  a Nelder-Mead approach \cite{nelder1965simplex} is utilized, where a Monte Carlo simulation technique \cite{kroese2013handbook} is employed  in each iteration step of the Nelder-Mead algorithm to approximate the expected value  of the loss $L\bigl(P_\theta\cap E,\restr{\G}{\aed(P_\theta)}\bigr)$.

This process involves averaging over numerous cross sections $P^{(i)}_\theta\cap E^{(i)}$, where $P^{(i)}_\theta$ is a realization of the extended crack network model $P_\theta$, and $E^{(i)}$ is a realization of the randomly orientated plane $E\subset\R^3$ for each $i=1,\ldots,n$ and some integer $n\in\N$. Recall that  $P_\theta$ is an isotropic model, i.e., the realizations of $P_\theta$ exhibit a statistically similar behavior in each direction. Thus, it would be sufficient, to intersect each realization $P^{(i)}_\theta$ of $P_\theta$ with a single plane $E_{\text{x},v}\subset\R^3$, where $E_{\text{x},v}$ denotes a plane that is orthogonal to the x-axis and has a certain distance $v>0$ from the origin $o\in\R^3$.  However, to keep the computational effort low and, simultaneously, increase the robustness of averaging, each realization $P^{(i)}_\theta$ is intersected at multiple distances along the x-, y- and z- axis, respectively.  Furthermore, to avoid interpolations of the pixelized image data, cross sections are only taken at integer heights along the coordinate axes.

First, 100 pristine particles are drawn from the stochastic 3D model for polycrystalline NMC particles, which has been described in Section~\ref{ssec:PristineModel}. Then, in each iteration step of the Nelder-Mead minimization algorithm, 32 out of these 100 particles, denoted by $P^{(i)}_\text{pr}=(\solid^{(\text{pr},i)},\emptyset)$ for $i=1,\ldots,32$, are chosen with a probability proportional to their volume-equivalent diameter. Note that this selection method corresponds to the probability of intersecting a particle by a randomly chosen plane, as this is done in 2D SEM imaging \cite{baddeley2004}.

Each  pristine particle $P^{(i)}_\text{pr}$ serves as input for generating a realization of the extended crack network model $P_\theta$, which results in 32 realizations of $P_\theta$, denoted by $P^{(i)}_\theta=(\solid^{(\theta,i)},\crack^{(\theta,i)})$ for $i=1,\ldots,32$. 
Additionally, for each realization  $P^{(i)}_\theta$, multiple cross sections  are generated by intersecting each simulated particle $P^{(i)}_\theta$ at $10\%,20\%,\ldots,90\%$ of its size along  the x-,y- and z-axis, respectively. This yields 32 realizations of $P_\theta$, each sliced at 9 positions along 3 axes, which finally results into $32 \times 9 \times 3 = 864$  cross sections per iteration step.

More formally, for each simulated particle $P_\theta^{(i)}$, we assume without loss of generality that it is located in the positive octant $\R_+^3=[0,\infty)^3\subset\R^3$ and touches the xy-plane, xz-plane and yz-plane. Furthermore, let 
$\diam_\text{x}(P_\theta^{(i)})$ denote the Feret diameter of $P_\theta^{(i)}$ \cite{merkus2009} along the x-axis, which is given by
\begin{align}\label{def.fer.dia}
    \diam_\text{x}(P_\theta^{(i)})=\max\Bigl\{v>0\colon \bigl(\solid^{(\theta,i)}\cup\crack^{(\theta,i)}\bigr)\cap E_{\text{x},v}\neq\emptyset\Bigr\},
\end{align}
i.e., $\diam_\text{x}(P_\theta^{(i)})$ describes the size of $P_\theta^{(i)}$ in x-direction. Analogously, the  Feret diameters of $P_\theta^{(i)})$ along the y- and z- axis will be denoted by $\diam_\text{y}(P_\theta^{(i)})$ and  $\diam_\text{z}(P_\theta^{(i)})$, where the plane $E_{\text{x},v}$ on the right-hand side of Eq.~\eqref{def.fer.dia} is replaced by  planes $E_{\text{y},v}$ and $ E_{\text{z},v}\in\R^3$ that are orthogonal to the y- and z-axis, respectively, and have the distance $v>0$ to the origin.

Then, the expected loss $\EX L\bigl(P_\theta\cap E,\restr{\G}{\aed(P_\theta)}\bigr)$, occurring in Eq. \eqref{eq:LossFunction}, is numerically approximated by
\begin{align*}
    \EX L\Bigl(P_{\theta}\cap E, \restr{\G}{\aed(P_{\theta}\cap E)}\Bigl) \approx
    \frac{1}{864} \sum_{i=1}^{32}\sum_{\text{a}\in\{\text{x,y,z}\}}\sum_{j=1}^{9}L
    \Bigl( 
    P_\theta^{(i)} \cap E(\text{a},j,P_\theta^{(i)}),
    \restr{\G}{\aed\bigl(
    P_\theta^{(i)} \cap E(\text{a},j,P_\theta^{(i)}),
    \bigr)}
    \Bigr)
\end{align*}
where $E(\text{a},j,P)=E_{\text{a},\text{round}(j/10\cdot\diam_{\text{a}}(P))}$
and round$(\cdot)$ denotes rounding to the closest integer, as defined in Eq.~\eqref{eq:defRoundToInt}.

Thus, in summary, to find the optimal  parameter vector $\widehat{\theta}$ which solves the minimization problem given in Eq.~\eqref{eq:LossFunction},    in each iteration step of the Nelder-Mead  algorithm we choose 32 pristine particles  out of a pool of 100 realizations of the stochastic 3D  model described in Section~\ref{ssec:PristineModel}. These selected particles serve as input for the extended crack network model $P_\theta$, where the expected  loss is approximated by averaging over 864 cross sections.

Recall that the optimization procedure described above was separately applied to both data sets, $\G\Short$ and $\G\Long$, resulting in two calibrated models which generate particles exhibiting predominately  short or long cracks. In the following, we will refer to the extended crack network model calibrated to $\G\Short$ and $\G\Long$ as $\modelShort$ and $\modelLong$, where samples drawn from these two models are called short- and long-cracked particles, respectively.

\section{Results and discussion}\label{sec:ResultsAndDiscussion}
In this section  we first show how   the extended  crack network model $P_{\widehat\theta}$ can be validated, the calibration of which to experimental data has been explained in Section~\ref{ssec:Fitting}. For this, further geometric descriptors of 2D morphologies will be introduced in Section~\ref{ssec:validationDescriptors}, which have not been used for model calibration. Then, similarly to the model calibration approach considered in Section \ref{ssec:Fitting}, the distributions of these   descriptors will be determined in Section~\ref{ssec:validation} for simulated 2D cross sections, drawn from  $P_{\widehat\theta}$, and  compared to those computed for experimental 2D SEM data.  
Moreover, in Section~\ref{sec:effectiveDescriptors}, two transport-relevant particle descriptors are presented, which influence the performance of Li-ion batteries, but can only be determined adequately if 3D data is available. In Section \ref{ssec:transportRelevantProperties}, the distributions of these transport-relevant descriptors, namely the relative shortest path length of Li transport in active material, as well as the relative specific surface area of active material, are analyzed for simulated (pristine and cracked) 3D particles.

\subsection{Additional geometric descriptors of 2D morphologies}\label{ssec:validationDescriptors}
For validating the goodness of model fit, six further descriptors of 2D morphologies are  taken into account to compare planar cross-sections of the extended  crack network model $P_{\widehat\theta}$ to experimentally measured 2D SEM images described in Section \ref{sec:Materials}. It is important to emphasize that the descriptors considered in the present section are not used during the calibration process explained in Section~\ref{ssec:Fitting}.  Furthermore, note that these descriptors are defined for planar particle cross-sections $P=(\solid,\crack)$, with $\solid,\crack\subset\R^2$, which are either the continuous representation of an experimentally measured particle cross-section, or derived by intersecting a simulated cracked 3D particle, drawn from $P_{\widehat\theta}$,  with a randomly oriented plane $E\subset\R^3$.

\vspace{3mm}
\paragraph{Porosity}
One of the most fundamental geometric descriptors of porous 2D morphologies is their porosity. In the case of a planar particle cross sections $P=(\solid,\crack)$, the porosity $p\in[0,1]$ can be given by
\begin{align*}
    p=\frac{\nu_2(\crack)}{\nu_2(\solid\cup\crack)},
\end{align*}
see also Eq.\eqref{eq:Porosity.neu} in Section \ref{ssec:CrackEnsembleModel}, where  the porosity was assumed to be independent of the particle size. However, recall that the porosity was not used  in Section~\ref{ssec:Fitting} for calibrating the extended crack network model  $P_{\widehat\theta}$ to experimental data.  In Section~\ref{ssec:validation}  we determine the (empirical) distribution of $p$ for simulated 2D cross sections, drawn from  $P_{\widehat\theta}$, and  compare it to that computed for experimental 2D SEM data.

\begin{figure}[h]
     \centering
     \includegraphics[width=.8\textwidth]{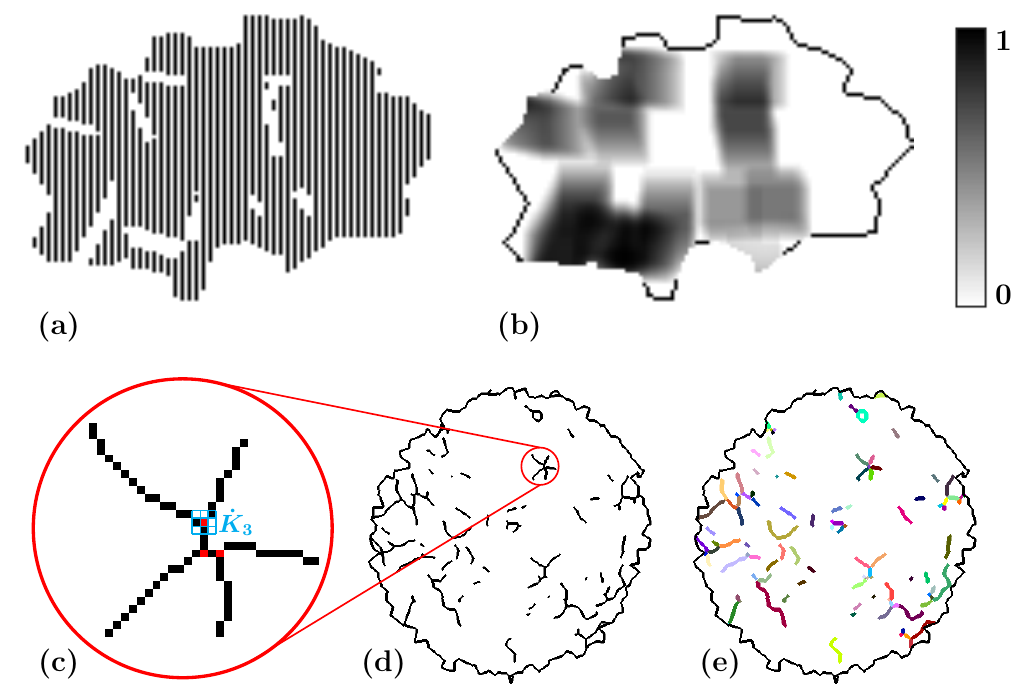}
     \caption{Geometric descriptors of 2D morphologies, including chord lengths (a) and local entropy (b) of an elongated particle, as well as the number of branching points within a magnified region (c) of a skeletonized crack network (d), along with the number and length of crack segments (e) in another, more spherical particle. Note that
     for illustrative purposes, the skeletons in (d) and (e) are dilated and 
     the grain boundaries in (b), (d) and (e) are indicated.
     }
     \label{fig:descriptors}
 \end{figure}

\vspace{3mm}
\paragraph{Chord length}
Let $v\in \{x\in\R^2:|x|=1\}$ be some predefined direction in $\R^2$. 
Then,  chords within the solid phase $\solid$ can be obtained by intersecting  $\solid$ with (parallel) lines in direction $v$. 
In general, this intersection results  in multiple line segments, which are referred to as chords, see Figure \ref{fig:descriptors}a for chords in y-direction, i.e., $v=(0,1)$.
The probability distribution of the lengths of these line segments is called chord length distribution. 
Under the assumption of isotropy, the chord length distribution does not depend on the chosen direction~$v$, see  \cite{chiu.2013} for
 formal definitions.
 
For 2D SEM data the chord length distribution was estimated by considering chords in x- and y-direction. However, for simulated 3D particles, drawn from  $P_{\widehat\theta}$, chords along the z-direction were additionally taken into account, which increases robustness of the estimation.
For the computation of chord lengths, the python package PoreSpy \cite{Gostick2019} was used.

\vspace{3mm}
\paragraph{Local entropy}
The  mean local entropy of a particle cross section $P=(\solid,\crack)$ is a measure for the local heterogeneity  of  $P$. It can be defined in the by the following: First, assign each point $x=(x_1,x_2)\in\solid\cup\crack\subset\R^2$  its local entropy 
\begin{align*}
E(x)=-\mkern-22mu \sum_
{\substack{\Xi\in \{\solid,    \crack\}}} 
\mkern-18mu\varepsilon_\Xi(x)\log_2\bigl(\varepsilon_\Xi(x)\bigr) 
\end{align*}
where $\varepsilon_\Xi(x)\in[0,1]$ denotes the local volume fraction of phase $\Xi\in\{\solid,\crack\}$, Note that $\varepsilon_\Xi(x)$ is determined by means of  the $15\times 15$ neighborhood $K_{15}(x)\subset\R^2$ centered iat $x\in\solid\cup\crack$, and formally given by 
\begin{align}\label{lec.ent.def}    \varepsilon_\Xi(x)=\frac{\nu_2\bigl(\Xi\cap K_{15}(x)\bigr)}{\nu_2\bigl(\solid\cup\crack\cap K_{15}(x)\bigr)},
\end{align}
where $K_{15}(x)=\{y=(y_1,y_2)\in\R^2\colon |x-y|\leq15\}$ with $|x-y|=|x_1-y_1|+|x_2-y_2|$, being the so-called  Manhattan metric. Then, the mean local entropy  of the particle cross section $P=(\solid,\crack)$ is given by 
\begin{align}\label{mea.loc.ent}
E(P)= \frac{1}{\nu_2(\solid\cup\crack)}\int_{\solid,\crack} E(x)\,\text{d}x  \,,
\end{align}
i.e., by averaging the local entropy $E(x)$ over all  $x\in\solid\cup\crack$ belonging either to the crack or  solid phase.

In Section~\ref{ssec:validation}, the distribution of the mean local entropy $E(P)$ given in Eq.~\eqref{mea.loc.ent}  will be estimated for 2D image data and, therefore, the local entropy $\varepsilon_\Xi(x)$ introduced in ~\eqref{lec.ent.def} will be determined pixelwise. However, note that the latter quantity is highly sensitive to  changes of  resolution, because a finer resolution corresponds to a kernel $K_{15}$ containing more pixels to cover a predefined area, potentially resulting in a higher local heterogeneity. Therefore, the experimental 2D SEM data were downsampled to match the (coarser) resolution of the virtual pristine particles drawn from the stochastic 3D model, as described in Section~\ref{ssec:PristineModel}. 
Figure~\ref{fig:descriptors}b illustrates a visual impression of local entropy computed on pixelized image data.

\vspace{3mm}
\paragraph{Number of branching points}
To investigate the branching behavior of the crack phase $\crack$ of a particle cross section $P=(\solid,\crack)$, we consider its skeleton, denoted by $\skel(P)$, see also Section~\ref{ssec:Subdivison}. Recall that in Section~\ref{ssec:Subdivison}, each connected component of the crack phase   has been represented by its center line, called skeleton segment and denoted by $S\in\skel(P)$, where the family of all skeleton segments  forms the skeleton. In the following, for each skeleton segment 
$S\in\skel(P)$, we say that $x\in S$ is  a branching point if there are at least three  points $y_1,y_2,y_3\in S$ such that  $|x-y_i|=\varepsilon$, where the set $\varepsilon>0$ is a sufficient small distance.

To estimate the distribution of the number of branching points from pixelized image data,  for any $S\in\skel(P)$ and $x\in S\cap\Z^2$, let $\dot{K}_3(x)=(K_3(x)\cap\Z^2)\setminus\{x\}$ denote the $3\times3$ neighborhood of $x$ on the grid $\Z^2$, excluding the point $x$ itself. A point $x\in S\cap\Z^2$ is considered a branching point if $\#(S\cap \dot{K}_3(x))\geq3$, see Figure~\ref{fig:descriptors}c, where $\dot{K}_3(x)$ is visualized in blue color.

The python package PlantCV has been used \cite{FAHLGREN2015} to compute skeleton segments and branching points.

\vspace{3mm}
\paragraph{Number and length of crack segments}
By removing the branching points from a skeleton segment $S\subset\skel(P)$, we obtain various connected components of $S$ which we refer to as crack segments, see Figure~\ref{fig:descriptors}e, where crack segments are indicated in different colors.
Furthermore, for  validation of the fitted extended crack network model $P_{\widehat\theta}$,
we determine the distributions of  the number and length of crack segments 
for simulated 2D cross sections, drawn from  $P_{\widehat\theta}$, and  compare them to those computed for experimental 2D SEM data. Note that the notion of  crack segment length introduced in this section is different from that of crack size, which was considered in Sections~\ref{sssec:descriptors}  to \ref{sssec:Training}
for model fitting.

\subsection{Model validation}\label{ssec:validation}

\begin{figure}[h]
    \centering
    \includegraphics[width=\textwidth]{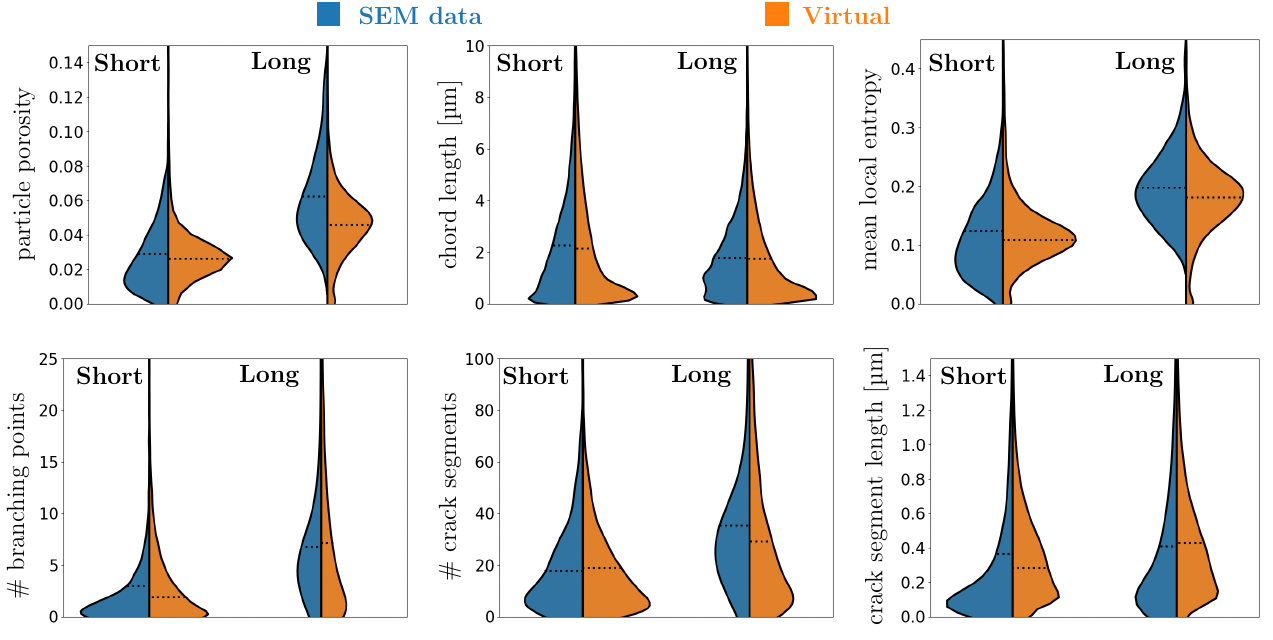}
    \caption{
    Probability densities of 
    porosity, chord lengths, mean local entropy (top row), number of branching points,  number and length of crack segments (bottom row). Blue areas indicate densities computed from SEM data, whereas orange areas correspond to  densities for planar cross sections of 3D realizations of the extended crack network model. Within each subplot, the left column corresponds to the data set $\G\Short$, and the right column to $\G\Long$. The horizontal dashed lines indicate the mean values of the respective descriptors. }
    \label{fig:ValidationViolines}
\end{figure}

To validate the extended  crack network model, which has been calibrated to experimental image data in  Sections~\ref{sssec:descriptors}  to \ref{sssec:Training}, the probability densities of the geometric descriptors stated in Section \ref{ssec:validationDescriptors} are estimated using particle cross sections of 200~model realizations drawn from each of the  extended crack models $\modelShort$ and $\modelLong$.  To ensure comparability, only 2D cross sections of the 3D realizations have been taken into account, which are extracted, similarly as described in Section~\ref{sssec:Training}, at $10\%,20\%,\ldots,90\%$ of the particle size along x-,y- and z-direction, resulting into $9\cdot3\cdot200=5400$ cross sections for both crack scenarios. For each of these cross sections, thee porosity, mean local entropy, number of branching points, as well as the number and length of crack segments are determined. Their probability densities, along with those derived from experimental 2D SEM data, have been computed via kernel density estimation, see Figure~\ref{fig:ValidationViolines}.

When comparing the probability densities shown in Figure \ref{fig:ValidationViolines},  derived for each case from simulated and experimental data,  respectively, it becomes clearly visible that these pairs of densities exhibit similar shapes, indicating a suitable  choice of model type  and a  quite good fit of model parameters, for both data sets $\G\Short$ and $\G\Long$. Even in cases where these pairs of probability densities are slightly different from each other, like the densities of the porosity of short-cracked particles (upper row, left part, left pair of  densities), their mean values, represented by horizontal dashed lines, fit very well. On the other hand, for example, the porosity distribution  of long-cracked particles (upper row, left part, right pair of densities)  exhibits a slightly larger deviation of its mean value with respect to the corresponding mean value derived from simulated data. Nevertheless, qualitatively, the overall shapes of the probability densities match  quite well in all cases.

In summary, the probability densities derived from simulated and experimentally measured image data show a high degree of agreement, indicating that the crack networks observed in 2D SEM data are accurately represented by the stochastic 3D model introduced in Section~\ref{sec:Model}. 

 \subsection{Transport-relevant particle descriptors in 3D}\label{sec:effectiveDescriptors}
In this section, two  geometric particle descriptors are considered, which influence the performance of Li-ion batteries, but can only be determined adequately if 3D image data is available. However, in general, the acquisition of  3D  data by tomographic imaging is expensive in terms of  time and costs. Therefore, in the present paper, these descriptors are estimated by means of a stochastic 3D model, i.e., from realizations of the extended crack network model $P_\theta$, which has been introduced in Section~\ref{sec:Model} and calibrated by means of 2D image data in Section~\ref{ssec:Fitting}.
 
Thus, in the following, we consider a cracked particle $P=(\solid,\crack)$, where $\solid,\crack\subset\R^3$, drawn from the extended crack network model $P_\theta$, %which was introduced in Section \ref{sec:Model}, 
and we consider its pristine counterpart $P_\text{pr}=(\solid^\prUp,\emptyset)$ with $\solid^\prUp=\solid\cup\crack\subset\R^3$, which serves as input for $P_\theta$. Furthermore, by $\BG=\R^3\setminus \solid^\prUp$ we denote the background of both particles, $P_\text{pr}$ and $P$.

In particular, relative shortest path lengths from  active material of $P$ to  electrolyte (located in cracks and/or  background)  are considered. Note that this  is an important  particle descriptor, since during delithiation, lithium ions migrate from the active material to the surface of the particle, where deintercalation  occurs.
Moreover, we investigate the specific surface area of  particles, showing how it is affected by cracking. Clearly, this is also a transport-relevant particle descriptor, because it characterizes the intercalating surface of a particle.

\vspace{3mm}
\paragraph{Relative shortest path lengths}\label{ssec:relativePathLengths}
The paths from randomly chosen locations within the active material to electrolyte are analyzed to investigate the transport of Li during delithiation. First, the case is considered that particles are embedded in liquid electrolyte, where open porosity cracks are filled with electrolyte and transport path lengths may decrease.  Furthermore, to demonstrate that the extended crack network model $P_\theta$ introduced in Section~\ref{sec:Model}  is not limited to Li-ion batteries with liquid electrolyte, the case of solid electrolyte is considered to mimic the behavior of all-solid-state batteries. Then, contrary to batteries with liquid electrolyte, cracks caused by cycling are not penetrated with electrolyte. Thus, cracks can be considered as obstacles to ion transport, which may increase transport path lengths.

A powerful tool to analyze transport paths within a given phase 
of a two-phase material is the so-called geodesic tortuosity. It is a purely geometric descriptor, see e.g. \cite{holzer2023}, which is usually  estimated on image data by considering two parallel planes in $\R^3$, the starting plane and the target plane, denoted by $E_\text{S}$ and $E_\text{T}$ in the following. Then, for each $x\in E_\text{S}$, the length of the shortest path   
to the target plane $E_\text{T}$ within the transport phase 
is determined and normalized by the distance between the planes $E_\text{S}$ and $E_\text{T}$. For estimating the geodesic tortuosity  in the formal framework of  random closed sets, we refer to \cite{neumann2019}. In the present paper, the concept of geodesic tortuosity is generalized by  considering arbitrary starting and target sets $\startingSet,\targetSet\subset\R^3$ such that $\startingSet\cap\targetSet=\emptyset$.

\begin{figure}[h]
    \centering
    \includegraphics[width=.99\textwidth]{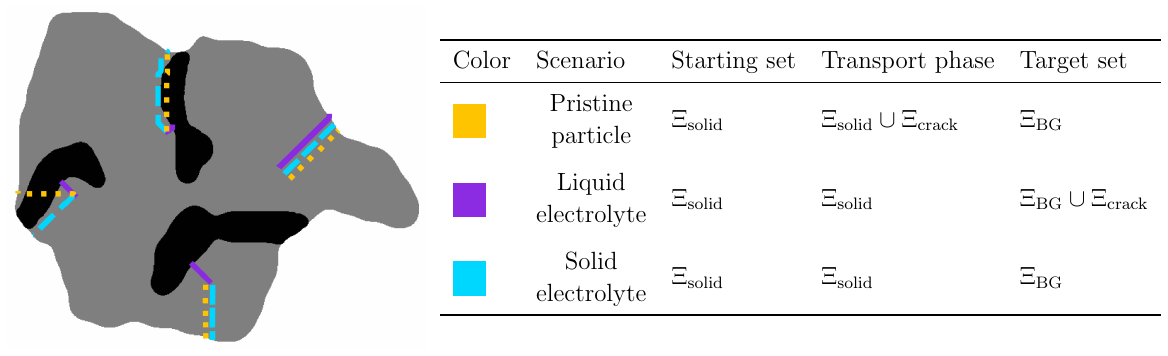}
    \caption{Shortest paths from  active material (grey) to electrolyte, avoiding cracks (black); for a  cracked particle  embedded in  liquid  (purple) and solid (blue) electrolyte, respectively,  and for the corresponding pristine particle (orange).}
    \label{fig:sketchShortestPath}
\end{figure}

To investigate delithiation in the case of liquid electrolyte (LE), the shortest paths from active material to electrolyte are determined by means of simulated 3D image data. For this, the starting and target sets $\startingSet,\targetSet$ are discretized, where  $\startingSet=\solid$ and $\targetSet=\crack\cup\BG$. 
Note that the union $\crack\cup\BG$ of cracks and background forms the continuous representation of the target set, since cracks are filled with liquid electrolyte.
On the other hand, to mimic the behavior of so-called all-solid-state batteries with solid electrolyte (SE), where cracks serve as obstacle, the  starting and target sets  
are given by  $\startingSet=\solid$ and $\targetSet=\BG$. 
In Figure~\ref{fig:sketchShortestPath}, examples of  shortest paths are shown for the cases of liquid and solid electrolyte, alongside with shortest paths in the corresponding pristine (i.e. non-cracked) particle.

For each  $x\in\startingSet$, the length  of the shortest path within the active material  from $x\in\startingSet$ to the target set $\targetSet$ is determined,  where the  transport phase is given by the set $\transportPhase=\solid$ for a cracked particle, and by $\transportPhase=\solid^\prUp=\solid\cup\crack$ for the corresponding pristine particle.
To compute these shortest path lengths, denoted by 
  $\gamma_\transportPhase(x,\targetSet)$,
 Dijkstra's algorithm \cite{diestel.2018} was utilized, as implemented in the python package dijkstra3D.

Moreover, to investigate how cracking affects the shortest path lengths, we consider relative shortest path lengths, denoted by $\LE(x,P)$ for liquid electrolyte and by $\SE(x,P)$ for solid electrolyte. These quantities are determined by normalizing 
the shortest path length $\gamma_\transportPhase(x,\targetSet)$, from  $x\in\startingSet$ to the target set $\targetSet$ within the transport phase $\transportPhase=\solid$ of a cracked particle, by the length of the corresponding shortest path within the solid phase $\solid^\prUp$ of the pristine particle $P_\text{pr}$.
Note that the shortest path within the pristine particle represents the shortest path from the  $x\in\startingSet$  to the electrolyte before the particle is cracked. 
Thus, formally, the relative shortest path lengths $\LE(x,P)$ and  $\SE(x,P)$ for liquid   and solid electrolyte, respectively, are as follows:
\begin{align}\label{rel.sho.pat}
    \LE(x,P)=\frac{
    \gamma_{\solid}\bigl(x,\crack\cup\BG\bigr)
    }{
    \gamma_{\solid^\prUp}\bigl(x,\BG\bigr)
    }\,, \qquad 
    \SE(x,P)=\frac{
    \gamma_{\solid}\bigl(x,\BG\bigr)
    }{
    \gamma_{\solid^\prUp}\bigl(x,\BG\bigr)
    }
\end{align}
for each $x\in  \startingSet=\solid$, where $\solid^\prUp=\solid\cup\crack$.

From Eq.~\eqref{rel.sho.pat} we get that  $\LE(x,P)\le 1$ and $\SE(x,P)\ge 1$ for each  $x\in  \startingSet=\solid$. This indicates a decrease of shortest path lengths caused by cracking in the case of liquid electrolyte, and an increase  for solid electrolyte, as expected. 

\begin{figure}[h]
    \centering
    \includegraphics[width=.75\textwidth]{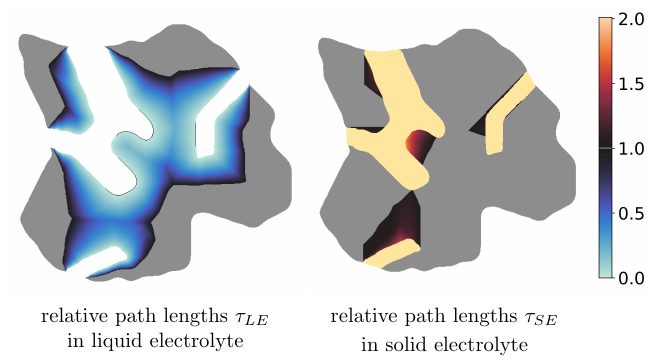}
    \caption{Relative shortest path lengths $\LE(x,P)$ and $\SE(x,P)$ for liquid (left) and solid electrolyte  (right). 
    Note that white indicates electrolyte, while bright yellow (right) indicate obstacles, formed by cracks. Additionally, gray within the particles corresponds to  relative path lengths equal to one,  indicating no change in the shortest path length due to cracking.}
    \label{fig:sketchTransportProperties}
\end{figure}

Finally, we consider the mean relative shortest path lengths $\LE(P)$ and $\SE(P)$, which are determined by averaging the relative shortest path lengths $\LE(x,P)$ and $\SE(P)$ given in  Eq.~\eqref{rel.sho.pat} over all $x\in\solid$. The concept of  relative shortest path lengths is visualized in Figure~\ref{fig:sketchTransportProperties} for both kinds of (liquid and solid) electrolyte.

\vspace{3mm}
\paragraph{Relative specific surface area} \label{ssec:specificSurfaceArea}
%\paragraph{Specific surface area}
Another descriptor related to effective properties of a particle $P=(\solid,\crack)$ is its specific surface area $\ssa(P)$. 
It indicates its  surface area per unit volume and is formally given by
\begin{align*}
    \ssa(P)=\frac{\mathcal{H}_2\bigl(\partial \solid\bigr)}{\nubig{\solid}},
\end{align*}
where $\mathcal{H}_2(\,\cdot\,)$ denotes the 2-dimensional Hausdorff measure, $\nu_3(\,\cdot\,)$ the 3-dimensional Lebesgue measure and $\partial \Xi$ the boundary of a set $\Xi$. 
Note that $\mathcal{H}_2(\,\cdot\,)$ measures the area of a 2-dimensional manifold and $\nu_3(\,\cdot\,)$ the volume of a 3-dimensional set. Recall that in the present paper model realizations are voxelized data. Therefore, the surface area of $\solid$ is estimated using the algorithm presented in \cite{ohser2009} and the volume by counting voxels associated with $\solid$.

%\paragraph{Relative specific surface area}
To investigate the change of the specific surface area  caused by cracking, the relative specific surface area, given by
\begin{align*}
    \ssa_\text{rel}(P)=\frac{\ssa(P)}{\ssa(P_\text{pr})}\,,
\end{align*}
is considered, where $P_\text{pr}$ denotes the underlying pristine particle corresponding to $P$. 
Note that the relative specific surface area $\ssa_\text{rel}(P)$ of $P$ quantifies the increase of surface area per unit volume due to cracking.  In particular,  $\ssa_\text{rel}(P)=1$  indicates no change, while larger values of $\ssa_\text{rel}(P)$  represent an increase in specific surface area caused by cracking. 
For example, $\ssa_\text{rel}(P)=2$  indicates a doubling of the specific surface area. Notably, in a real Li-ion battery system, the increase in specific surface area due to cracking is only beneficial for liquid electrolyte systems.  Additionally, the relative activity of newly exposed surfaces to electrochemical reactions will depend on the availability of an electron at the surface between the electrolyte and active material phase, which is not considered in the present work.

\subsection{Structural analysis of simulated 3D particles}
\label{ssec:transportRelevantProperties}
We now deploy the stochastic 3D model $P_{\widehat\theta}$ of cracked particles that has been calibrated by means of 2D data to investigate the transport-relevant descriptors stated in Section~\ref{sec:effectiveDescriptors} for simulated 3D particles drawn from$P_{\widehat\theta}$. In particular, we  investigate the probability distributions of the (relative) specific surface area and  the mean relative shortest path length (for solid and liquid electrolyte) associated with the stochastic 3D model $P_{\widehat\theta}$.
More precisely, we will provide a detailed discussion of the corresponding probability densities of these descriptors, separately for the stochastic 3D  model $\modelShort$ calibrated to the data set $\G\Short$, and for $\modelLong$ calibrated to $\G\Long$.

\begin{figure}[h]
    \centering
    \includegraphics[width=.7\textwidth]{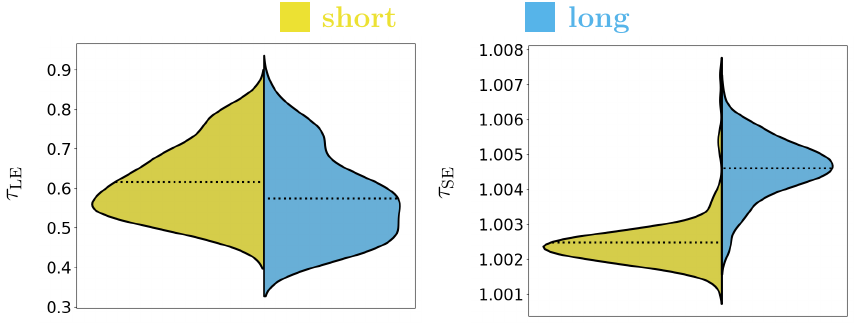}
    \caption{
    Probability densities of mean relative shortest path lengths  $\LE(P)$  and $\SE(P)$ for liquid (left) and solid electrolyte  (right).
    Each subfigure shows two probability densities, where the green (left) areas correspond to the probability densities computed from short-cracked particles and the blue (right) areas indicate the probability densities derived from long-cracked particles.} \label{fig:ResultsTransportProperties}
\end{figure}

\begin{figure}[h]
    \centering
    \includegraphics[width=0.7\textwidth]{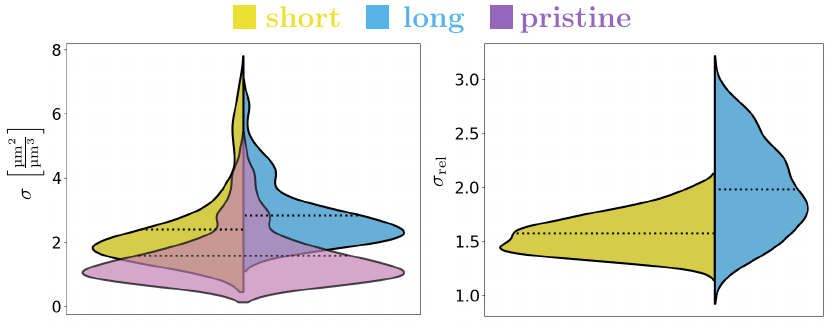}
    \caption{Left: Probability densities of the specific surface area $\ssa(P)$ for pristine particles (purple) and  for cracked particles  drawn from the stochastic 3D models  $\modelShort$ (green) and $\modelLong$ (blue). Right: Probability densities of the relative specific surface area
 $\ssa_\text{rel}(P)$ for cracked particles  drawn from  $\modelShort$ (green) and $\modelLong$ (blue), respectively.
       }
    \label{fig:ResultsEffectiveSSa}
\end{figure}

First, we draw 200  realizations from  $\modelShort$ which we denote by $P^{(i)}$ for $i=1,\dots,200$. By computing the transport-relevant descriptors introduced in Section~\ref{sec:effectiveDescriptors} for these realizations, we obtain four sample data sets, denoted by $\{\LE(P^{(i)})\}_{i=1}^{200}$, $ \{\SE(P^{(i)})\}_{i=1}^{200} $, $ \{\ssa(P^{(i)})\}_{i=1}^{200}$ and $\{\ssa_\text{rel}(P^{(i)})\}_{i=1}^{200}$.    Then, by means of kernel density estimation on each of these four sets, we get probability densities of the corresponding transport-relevant particle descriptors, see the blue plots in Figures~\ref{fig:ResultsTransportProperties} and \ref{fig:ResultsEffectiveSSa}. Furthermore, the same procedure was applied to 200 realizations drawn from $\modelLong$ to determine probability densities  of the particle descriptors introduced in Section~\ref{sec:effectiveDescriptors}, see the green plots in Figures~\ref{fig:ResultsTransportProperties} and \ref{fig:ResultsEffectiveSSa}.

Note that the transport-relevant particle descriptors introduced in Section~\ref{sec:effectiveDescriptors}, with the  exception of the specific surface area $\ssa(P)$, are computed by comparing descriptors of simulated cracked particles with those of the underlying  pristine counterparts. Consequently, for pristine particles the mean relative shortest path length as well as the relative specific surface area are deterministic (i.e. non-random) quantities,  being equal to $1$. Therefore, when considering probability distributions of relative transport-relevant descriptors, only the specific surface area of pristine particles, see Figure~\ref{fig:ResultsEffectiveSSa} (left, purple), is of further interest.

The comparison of the probability densities shown in Figures~\ref{fig:ResultsTransportProperties} and \ref{fig:ResultsEffectiveSSa}
provides us with quantitative insight into the transport behavior of cracked 3D particles, even though initially only 2D data was available.
 For example, an intuitive result is that  shortest path lengths decrease after cracking for liquid electrolyte systems, i.e., the mean relative shortest path lengths are typically smaller than 1, see Figure \ref{fig:ResultsTransportProperties} (left). This is to be expected as cracks can be flooded by the liquid electrolyte leading to shorter transport paths. 
On the other hand, shortest path lengths  increase for  solid electrolyte, even though the relative increase is  marginal, i.e., only slightly above 1, see
Figure \ref{fig:ResultsTransportProperties} (right).

These general trends can be observed for both variants of the calibrated stochastic 3D  model, $\modelShort$ and  $\modelLong$. However, when comparing both models, we observe that---in the case of liquid electrolyte---mean shortest path lengths seem to decrease more significantly for  long-cracked particles rather than for short-cracked ones. For solid electrolyte systems, the difference in mean shortest path lengths between short- and long-cracked particles is much smaller, taking into account the finer length scale  of the y-axis on the right-hand side of Figure \ref{fig:ResultsTransportProperties}.

In the case of  liquid electrolyte, an explanation for the existence of shorter transport paths is the fact that transport paths, which are originating in the active material phase, have the option to end at the interface between  active material and  crack phases, instead of ending at the background. In other words, caused by cracking, the set of possible endpoints of transport paths originating in the active material becomes larger which possibly leads to a decrease  of shortest path lengths.  
On the other hand, in the case of  solid electrolyte, only a small fraction of shortest transport paths seems to be affected  by  cracking (which can cause obstacles to form). Consequently, we observe mean relative shortest path lengths close to 1, see Figure \ref{fig:ResultsTransportProperties} (right). 
From  the 2D illustrations of relative shortest path lengths in liquid and solid electrolyte, shown in Figure~\ref{fig:sketchTransportProperties}, a visual impression of this effect can be obtained.

With respect to specific surface area, see Figure \ref{fig:ResultsEffectiveSSa} (left), we observe that both scenarios (i.e., short- and long-cracked particles) lead to an increase of this geometric particle descriptor---an effect that is more pronounced for long-cracked particles generated by $\modelLong$. 
Moreover, the relative specific surface area quantifies this increase  compared to the underlying pristine particle. Short-cracked particles exhibit an average increase in their specific surface area  by a factor of 1.5 in comparison to their pristine counterparts, whereas this factor is equal to 2 for long-cracked particles, see Figure~\ref{fig:ResultsEffectiveSSa} (right).

In summary, it is important to note that the transport-relevant particle descriptors discussed in this section, namely the mean relative shortest path length and the relative specific surface area of cracked  particles, are just examples of  numerous further descriptors of 3D particles, which cannot be adequately determined from 2D cross sections.  Thus, the stereological  approach to stochastic 3D modeling of cracked particles proposed in the present paper can be used in future research to provide geometry input for spatially resolved numerical modeling and simulation, with the goal to derive quantitative structure-property relationships of cathode materials in Li-ion batteries, e.g. with respect to mechanical and electrochemical properties.

\section{Conclusion}

This paper presents a novel approach for generating virtual 3D cathode particles with crack networks that are statistically equivalent to those observed in 2D cross-sections of experimentally manufactured particles, where a stochastic 3D model is developed which inserts cracks into virtually generated NMC particles, requiring solely 2D image data for model calibration.

An essential advantage of our model is that it enables the generation and analysis of a large number of virtual  particle morphologies in 3D, whose planar 2D sections exhibit similar statistics as planar sections experimentally manufactured particles. %  fitted to a large number of 2D cross sections. 
This computer-based procedure is cheaper, faster and more reliable than analyzing just a few experimentally  manufactured particles. One reason for this is the circumstance that the acquisition of tomographic  image data  for a statistically representative number of particles can be expensive in both time and resources. 

On the other hand, virtual particles generated by our stochastic 3D model  allow for a more rigorous quantification of cracked NMC particles, i.e., by characterizing their 3D morphology and, subsequently,  by conducting spatially resolved mechanical and electrochemical simulations. This  supports the  analysis and comparison of different cycling conditions such as varying C-rate, operating temperature, or number of cycles.

It is important to emphasize that the stochastic model presented in this paper for the 3D morphology of cracked NMC particles is characterized by a small number of (nine) interpretable parameters. In contrast to  convolutional neural networks (CNNs), which have tens of thousands to several million trainable parameters, our stochastic 3D model has no \enquote{black-box} behavior and represents a low-parametric, transparent alternative to CNNs.

Moreover, our stochastic 3D model can be modified to involve further features that might influence cracking, e.g., by generating cracks  in dependence of the crystallographic orientation of  adjacent intraparticular grains.
For example, this can be achieved by considering  the misorientation between two neighboring grains, either replacing or supplementing the spatial alignment of the joint grain boundary. To implement such a modified model, orientation data of NMC particles  is required, which could be derived, e.g., from  EBSD measurements.

Another advantage of our stereological modeling approach is the fact that it allows for the estimation of chemo-mechanical properties from 2D images. More precisely, since our model only requires 2D images to generate realistic 3D particle  morphologies, it is possible to use these 3D morpohologies as geometry input for spatially resolved simulations of effective particle properties, which would be otherwise impossible to get them on the basis of  2D image data.

\section*{Acknowledgement}
 
This work was authored in-part by Alliance for Sustainable Energy, LLC, the manager and operator of the National Renewable Energy Laboratory for the U.S. Department of Energy (DOE) under Contract No. DE-AC36-08GO28308. Funding was provided by DOE’s Vehicle Technologies Office, Extreme Fast Charge and Cell Evaluation of Lithium-ion Batteries Program, Jake Herb, Technology Manager. The views expressed in the article do not necessarily represent the views of the DOE or the U.S. Government.

\bibliographystyle{myAbbrvnat}

\end{document}